\documentclass[aps,pra,twocolumn,superscriptaddress,longbibliography,floatfix]{revtex4-1}

\usepackage{amsmath,amssymb,graphicx,bm,dsfont}
\usepackage{xcolor}
\usepackage[colorlinks=true,urlcolor=blue,linkcolor=blue,citecolor=blue,bookmarks=false]{hyperref}

\begin{document}

\renewcommand{\vec}[1]{\boldsymbol{#1}}


\title{Symmetry breaking, slow relaxation dynamics, and topological defects\\at the field-induced helix reorientation in MnSi}

\author{A.~Bauer}
\email{andreas.bauer@frm2.tum.de}
\affiliation{Physik-Department, Technische Universit\"{a}t M\"{u}nchen, D-85748 Garching, Germany}

\author{A.~Chacon}
\affiliation{Physik-Department, Technische Universit\"{a}t M\"{u}nchen, D-85748 Garching, Germany}

\author{M.~Wagner}
\affiliation{Physik-Department, Technische Universit\"{a}t M\"{u}nchen, D-85748 Garching, Germany}

\author{M.~Halder}
\affiliation{Physik-Department, Technische Universit\"{a}t M\"{u}nchen, D-85748 Garching, Germany}

\author{R.~Georgii}
\affiliation{Heinz Maier-Leibnitz Zentrum (MLZ), Technische Universit\"{a}t M\"{u}nchen, D-85748 Garching, Germany}
\affiliation{Physik-Department, Technische Universit\"{a}t M\"{u}nchen, D-85748 Garching, Germany}

\author{A.~Rosch}
\affiliation{Institute for Theoretical Physics, Universit\"{a}t zu K\"{o}ln, D-50937 K\"{o}ln, Germany}

\author{C.~Pfleiderer}
\affiliation{Physik-Department, Technische Universit\"{a}t M\"{u}nchen, D-85748 Garching, Germany}

\author{M.~Garst}
\affiliation{Institute for Theoretical Physics, Universit\"{a}t zu K\"{o}ln, D-50937 K\"{o}ln, Germany}
\affiliation{Institut f\"{u}r Theoretische Physik, Technische Universit\"{a}t Dresden, D-01062 Dresden, Germany}

\date{\today}

\begin{abstract}
We report a study of the reorientation of the helimagnetic order in the archetypal cubic chiral magnet MnSi as a function of magnetic field direction. The reorientation process as inferred from small-angle neutron scattering, the magnetization, and the ac susceptibility is in excellent agreement with an effective mean-field theory taking into account the precise symmetries of the crystallographic space group. Depending on the field and temperature history and the direction of the field with respect to the crystalline axes, the helix reorientation may exhibit a crossover, a first-order, or a second-order transition. The magnetization and ac susceptibility provide evidence that the reorientation of helimagnetic domains is associated with large relaxation times exceeding seconds. At the second-order transitions residual Ising symmetries are spontaneously broken at continuous elastic instabilities of the helimagnetic order. In addition, on the time scales explored in our experiments  these transitions are hysteretic as a function of field suggesting, within the same theoretical framework, the formation of an abundance of plastic deformations of the helical spin order. These deformations comprise topologically non-trivial disclinations, promising novel routes to spintronics applications alongside skyrmions discovered recently in the same class of materials.
\end{abstract}

\vskip2pc

\maketitle


\section{Introduction}

\begin{figure}
\includegraphics[width=1.0\linewidth]{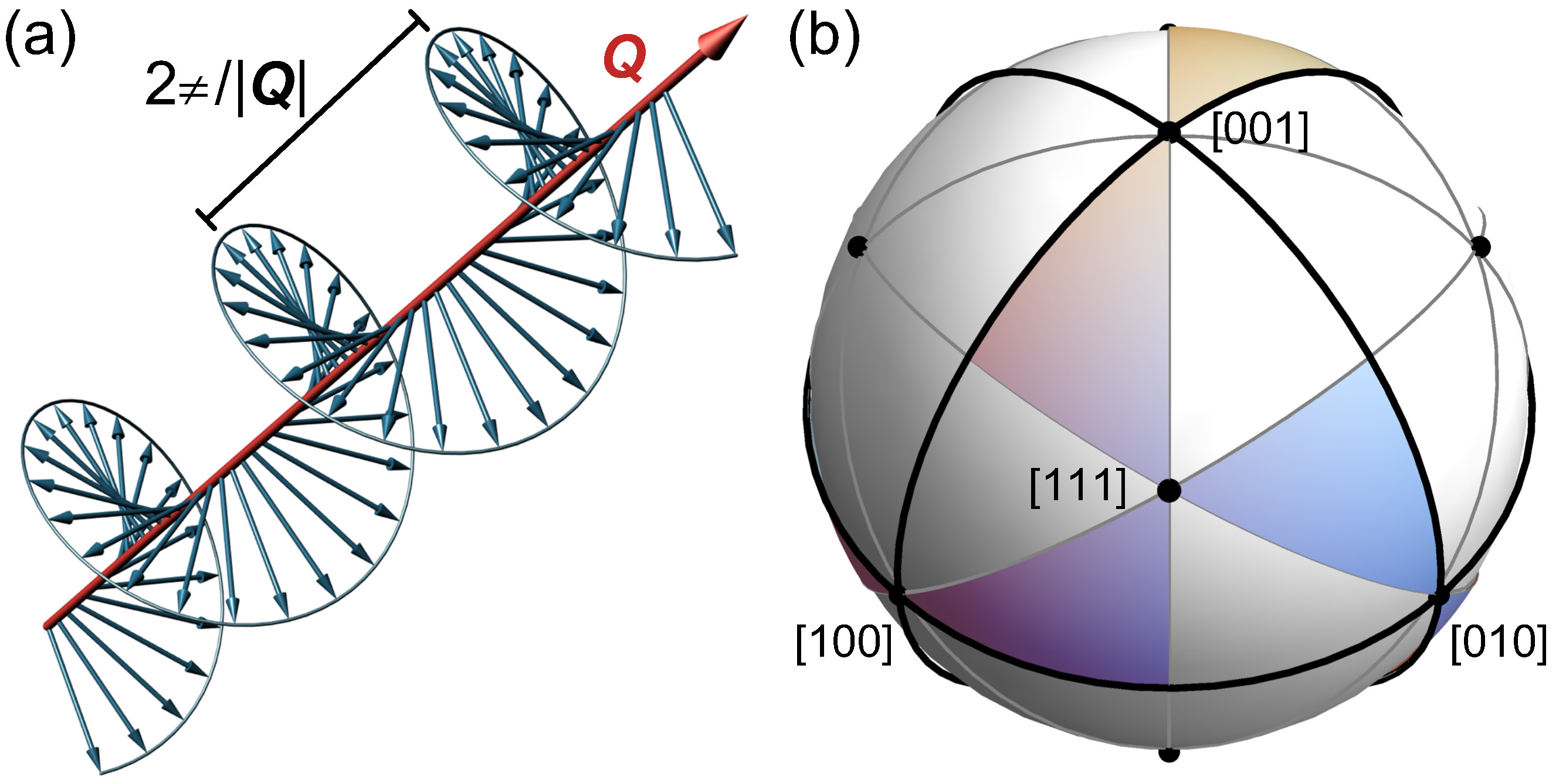}
\caption{Helimagnetism in MnSi. (a)~Magnetic helix with pitch vector $\vec{Q}$ (red arrow). (b)~Tetrahedral point group $T$ symmetry of MnSi with two-fold rotation axes along $\langle100\rangle$ and three-fold rotation axes along $\langle111\rangle$. The black great circles separate octants of the sphere centered around $\langle111\rangle$.}
\label{figure1}
\end{figure}

The characteristics of the temperature versus magnetic field phase diagram of magnetic materials, albeit frequently very subtle, reflect directly the full details of the underlying material-specific interactions. In recent years, compounds in which the ordered moments stabilize spontaneously a helix with a preferred chirality and wavevector $\vec{Q}$, illustrated in Fig.~\ref{figure1}(a), have been generating great research activities. This scientific interest experienced a major boost with the discovery that the application of magnetic fields may lead to the formation of soliton lattices~\cite{2012:Togawa:PhysRevLett, 2015:Kishine:Book} or topologically non-trivial forms of magnetic order, such as skyrmion or monopole lattices~\cite{2009:Muhlbauer:Science, 2009:Neubauer:PhysRevLett, 2010:Yu:Nature, 2011:Adams:PhysRevLett, 2011:Yu:NatureMater, 2012:Seki:Science, 2013:Milde:Science, 2015:Tanigaki:NanoLett, 2015:Tokunaga:NatCommun}. 

While these new forms of order are being explored intensely, an important unresolved question addressed in the following concerns the initial response of the helimagnetic state under small applied magnetic fields. In this limit, the precise symmetries of the magnetic anisotropies permitted by the crystallographic space group are decisive. Consequently, the accurate account of the magnetic phase diagram represents an important point of reference for emergent phenomena in condensed matter magnetism including the formation of any novel phases.

Two different limits may be distinguished regarding the initial response of the magnetic helix to an applied magnetic field as exemplified by hexagonal Cr$_{1/3}$NbS$_{2}$~\cite{2012:Togawa:PhysRevLett, 2015:Kishine:Book} and the cubic chiral magnets crystallizing in the non-centrosymmetric space group $P2_{1}3$ such as MnSi~\cite{1975:Bloch:PhysLettA, 1981:Plumer:JPhysCSolidStatePhys, 1981:Kataoka:JPhysSocJpn, 1989:Walker:PhysRevB, 2006:Grigoriev:PhysRevB2}, FeGe~\cite{1989:Lebech:JPhysCondensMatter}, Fe$_{1-x}$Co$_{x}$Si~\cite{2007:Grigoriev:PhysRevB}, or Cu$_{2}$OSeO$_{3}$~\cite{2012:Adams:PhysRevLett}.
In Cr$_{1/3}$NbS$_{2}$ relatively strong anisotropies fix the helix wavevector $\vec{Q}$ to the crystallographic $c$ axis and a magnetic field $\vec{H}\perp\vec{Q}$, starting from zero, gradually deforms the helix by partially polarizing the magnetic moments to become a well-understood incommensurate chiral soliton lattice~\cite{2012:Togawa:PhysRevLett, 2015:Kishine:Book}. Increasing the field further eventually results in a single transition to a field-polarized phase as predicted in a seminal paper by Dzyaloshinskii~\cite{1964:Dzyaloshinskii:SovPhysJETP}. 

In contrast, small magnetic fields are already sufficient to overcome very weak magneto-crystalline anisotropies as for the cubic chiral helimagnets, causing a reorientation of the entire helix at a characteristic field $H_{c1}$ into the so-called conical state for which $\vec{Q}\parallel\vec{H}$. Similar to the spin-flop transition in conventional antiferromagnets, the magnetic moments in the conical state gain Zeeman energy by canting towards the field without compromising the periodicity of the helix. When increasing the field further, the angle enclosed by the moments and the helix axis $\vec{Q}$ monotonically decreases until the helix amplitude vanishes in a second-order XY transition~\cite{2013:Bauer:PhysRevLett} at a critical field $H_{c2} > H_{c1}$. 

In this paper, we report a combined experimental and theoretical study of the rich phenomena associated with the reorientation process of the helix at $H_{c1}$ in MnSi. Experimentally we infer the precise nature and dynamics of the reorientation of the helical modulation from small-angle neutron scattering, magnetization, and ac susceptibility measurements for different field orientations. The experimental results are compared with the theoretical predictions deriving from an effective potential of the helix vector $\vec{Q}$ as determined by the symmetries of the tetrahedral point group $T$ of MnSi, see Fig.~\ref{figure1}(b). 

The excellent agreement between experiment and theory establishes a transparent and tractable starting point of the physical nature of the reorientation. In particular, our analysis of the magneto-crystalline anisotropy potential reveals that the reorientation for general magnetic field directions represents a crossover phenomenon or involves continuous (second-order) transitions for certain high-symmetry directions. Namely, for field directions corresponding to the great circles connecting two $\langle100\rangle$ directions on the unit sphere (black lines in Fig.~\ref{figure1}(b)), a residual Ising $\mathds{Z}_{2}$ symmetry is broken spontaneously. In this context a special situation arises for fields along the principle $\langle100\rangle$ axes where a $\mathds{Z}_{2}\times\mathds{Z}_{2}$ symmetry is spontaneously broken at two subsequent transitions, verifying the predictions of Walker~\cite{1989:Walker:PhysRevB}. Further, after zero-field cooling energetically unfavorable domains are depopulated in discontinuous first-order transitions. In turn, the excellent agreement between theory and our susceptibility and neutron scattering data permits the quantitative determination of the parameters specifying the magneto-crystalline potential. 

Taken together, these aspects reveal that the transitions of the helix orientation differ distinctly from conventional magnetic transitions. Notably, for large magnetic domains even a slight change of the helix axis involves a macroscopic reconstruction of the magnetization, implicating at least four main consequences as follows: (i)~Whereas the transitions in MnSi are described by an Ising order parameter, they do not belong to the Ising universality class. Instead, they are similar to elastic transitions in atomic crystals~\cite{2015:Zacharias:EurPhysJSpecialTopics, 2015:Zacharias:PhysRevLett}. (ii)~The reorientation of large magnetic domains occurs on macroscopic rather than microscopic time scales, which is confirmed by our susceptibility data showing two relaxation processes with well-separated time scales. Single magnetic moments relax essentially instantaneously, while the relaxation of the orientation of $\vec{Q}$ is slow and involves time scales exceeding seconds. (iii)~We find hysteretic behavior at transitions that are nominally continuous. Whereas sharp signatures characteristic of critical behavior are observed when the transition is approached from a single-domain state, these signatures are smoothed out when the transition is instead approached from a multi-domain state. We believe that this finding is associated with (iv)~the presence of topological defects which are, in particular, expected at domain boundaries. Frozen-in configurations of these defects may prohibit the equilibration of the system close to a reorientation transition giving rise to the hysteresis observed.

Our paper is organized as follows. After a detailed account of previous work in chiral helimagnets and the broader context in Sec.~\ref{Previous}, an overview of the experimental methods is given in Sec.~\ref{ExpMeth}, as well as a detailed account of the theoretical framework in Sec.~\ref{Theo}. Here we derive an effective theory for the helix vector $\vec{Q}$ emphasizing the striking simplicity encountered in cubic chiral magnets in the limit of weak magneto-crystalline anisotropies. A comparison with experimental data in Sec.~\ref{ExpRes} allows to determine the parameters of this model quantitatively. In Sec.~\ref{SANS}, we use small-angle neutron scattering for magnetic fields along $\langle110\rangle$ and $\langle100\rangle$ to track the evolution of the helix vector $\vec{Q}$ microscopically. Section~\ref{MagSus} is dedicated to measurements of the susceptibility. We address the response of the magnetic system on different time scales and show that the behavior after zero-field cooling may be reproduced by including finite-temperature effects. In addition, data is presented for a large number of different field directions. The paper concludes in Sec.~\ref{Discuss} with a discussion of our results.


\section{Further Motivation}
\label{Previous}

The class of cubic chiral magnets has been of great interest for many decades. Following the seminal work of Dzyaloshinskii~\cite{1957:Dzyaloshinskii:SovPhysJETP} and Moriya~\cite{1960:Moriya:PhysRevLett}, the identification of helimagnetism in MnSi and related compounds provided an important milestone in studies of complex modulated forms of order in the 1980s. Back then, it was the first example of an incommensurate long-wavelength modulation of an ordered state driven by the Dzyaloshinskii-Moriya spin-orbit interaction~\cite{1976:Ishikawa:SolidStateCommun, 1976:Motoya:SolidStateCommun, 1980:Bak:JPhysCSolidState, 1980:Nakanishi:SolidStateCommun}. Further, starting in the 1980s, studies of the spin fluctuation spectra, electronic structure, and magnetic equation of state in MnSi, which ignored the effects of spin-orbit coupling and weak anisotropies, provided a major breakthrough for itinerant electron ferromagnetism and established the starting point of studies of quantum phase transitions~\cite{1985:Lonzarich:JPhysCSolidState, 1985:Moriya:Book}. Finally, as the most recent development, the discovery of a skyrmion lattice in a small phase pocket in finite magnetic fields has attracted great interest~\cite{2009:Muhlbauer:Science, 2009:Neubauer:PhysRevLett, 2010:Yu:Nature}. For all of these different properties, a detailed understanding of the role of weak magnetic anisotropies is of great relevance, encompassing issues related to (i)~the incommensurability of the helical state at zero field, (ii)~itinerant-electron magnetism and the enigmatic non-Fermi liquid behavior reported at the quantum phase transitions~\cite{2001:Pfleiderer:Nature, 2007:Pfleiderer:Science, 2013:Ritz:Nature}, and (iii)~the formation of the skyrmion lattice phase~\cite{2009:Muhlbauer:Science} as anticipated in early theoretical work~\cite{1989:Bogdanov:SovPhysJETP, 2006:Roessler:Nature, 2010:Butenko:PhysRevB}. 

The first experimental observation of the reorientation of helimagnetic order in MnSi at a small field $H_{c1}$ was already achieved by Ishikawa \textit{et al.}~\cite{1976:Ishikawa:SolidStateCommun, 1982:Ishikawa:PhysRevB} with the help of small-angle neutron scattering, including first evidence on domain repopulations. Lebech \textit{et al.}~\cite{1993:Lebech:Book, 1989:Lebech:JPhysCondensMatter}, Grigoriev \textit{et al.}~\cite{2007:Grigoriev:PhysRevB}, and Adams \textit{et al.}~\cite{2012:Adams:PhysRevLett} subsequently reported similar behavior for the magnetic helix in FeGe, Fe$_{1-x}$Co$_{x}$Si, and Cu$_{2}$OSeO$_{3}$, respectively. In the magnetization, the helix reorientation results in a non-linear dependence on the applied field which was detected in early work on MnSi~\cite{1975:Bloch:PhysLettA, 1977:Hansen:PhD}, see Ref.~\cite{2010:Bauer:PhysRevB} for a recent study. 
This previous experimental work focused on a few crystallographic high-symmetry directions and it also did not investigate the transitions in sufficient detail.

On the theoretical side, Plumer and Walker \cite{1981:Plumer:JPhysCSolidStatePhys} as well as Kataoka and Nakanishi \cite{1981:Kataoka:JPhysSocJpn} first addressed the helix reorientation in MnSi on the level of the Ginzburg-Landau theory for the magnetization of Refs.~\cite{1980:Bak:JPhysCSolidState, 1980:Nakanishi:SolidStateCommun}. These studies identified the competition of the Zeeman energy and magneto-crystalline anisotropies to be at its origin. In particular, Plumer and Walker predicted a second-order transition for field orientations $\langle110\rangle$ and $\langle100\rangle$ with a ratio of critical fields $H^{\langle100\rangle}_{c1}/H^{\langle110\rangle}_{c1} \approx \sqrt{2}$. Subsequently, it was pointed out by Walker~\cite{1989:Walker:PhysRevB} that for fields along $\langle100\rangle$ the low symmetry of the space group should actually result in a splitting of the single transition into two, which so far had not been experimentally verified. The results of this early work, however, did not quantitatively describe the magnetization curve because the magnetic susceptibility transverse to the helix axis, $\chi_{\perp}$, was not computed exactly. 
More recently, Grigoriev \textit{et al.}~\cite{2006:Grigoriev:PhysRevB, 2007:Grigoriev:PhysRevB} suggested a transverse susceptibility $\chi_{\perp}$ that is correct in the limit $H \to 0$, but they did not exploit it for a full analysis of the helix reorientation process.

Whereas general aspects of the helix reorientation are known, important open questions concern the experimentally observed properties of the helix reorientation under the symmetry constraints of the non-centrosymmetric space group $P2_{1}3$ for arbitrary field directions. Related to this issue, a key theoretical question is whether the details of the reorientation of the helix as a function of field direction may be captured in a single tractable mean-field model. Further, the nature of the reorientation process and the nature of repopulation of helimagnetic domains were unresolved prior to our study, also alluding to the origin of the characteristic time-scales as observed by different experimental probes. In this context, questions arise on general similarities compared to other phase transitions, such as elastic transformations of crystal lattices, on the relevance of the underlying symmetries, and on the potential existence and character of defects of the order parameter. Last but not least, in view of the soliton lattice observed in Cr$_{1/3}$NbS$_{2}$, an obvious concern is the possible formation of solitonic modulations in cubic chiral magnets and to what extend the harmonicity of the helical modulation may get lost under applied magnetic field.


\section{Experimental Methods}
\label{ExpMeth}

For our study, single crystals of MnSi were grown by means of optical float-zoning under ultra-high vacuum compatible conditions~\cite{2011:Neubauer:RevSciInstrum, 2016:Bauer:RevSciInstrum, 2016:Reiner:SciRep}. The residual resistivity ratio of samples from these crystals is around 80, i.e., a typical value reported in the literature. From the single-crystal ingots we prepared three samples. Sample \#1 is a sphere with a diameter of 5.75~mm. Samples \#2 and \#3 are cubes with an edge length of 2~mm and surfaces perpendicular to $[110]$, $[1\bar{1}0]$, $[001]$ and $[110]$, $[1\bar{1}1]$, $[\bar{1}12]$, respectively. The samples were oriented using X-ray Laue diffraction. Spheres as well as cubes for field along their edges~\cite{1998:Aharoni:JApplPhys} exhibit a demagnetization factor $N = 1/3$ allowing to readily compare data for the samples geometries used in this study. All given field values are values of the applied magnetic field. The spherical geometry of sample \#1 further minimizes potential complexities arising from inhomogeneities of the internal magnetic fields due to inhomogeneous demagnetizing effects.

Small-angle neutron scattering was carried out on sample \#1 using the diffractometer MIRA2 at the Heinz Maier-Leibnitz Zentrum (MLZ) at an incident neutron wavelength of $(4.5\pm0.5)~\textrm{\AA}$~\cite{2015:Georgii:JLSFR}. The sample resided in a closed-cycle cryostat and a bespoke pair of Helmholtz coils allowed to apply a magnetic field perpendicular to the incoming neutron beam. A rotatable sample stick permitted to rotate the sample by $360^{\circ}$ around the field axis. For further details of the neutron scattering setup, the analysis of the data, and the construction of the spheres shown in Sec.~\ref{SANS}, we refer to the supplementary information~\cite{supplement}.

On samples \#2 and \#3 we measured the magnetization and the ac susceptibility at an excitation frequency of 911~Hz and with an excitation amplitude of 1~mT in a Quantum Design physical property measurement system. On the spherical sample \#1 magnetization was measured using an Oxford Instruments vibrating sample magnetometer and a bespoke sample holder that permitted to rotate the sample around a crystalline $\langle110\rangle$ axis. The angle between a $\langle100\rangle$ axis perpendicular to the latter and the magnetic field direction was determined with an optical microscope, where the total uncertainty of the sample orientation is estimated to be $\pm1^{\circ}$. The field values were changed in 1~mT steps and subsequently the magnetization was detected by integrating the oscillations at 62.35~Hz over 3~s while keeping the field constant. The susceptibility was calculated by numerically differentiating the measured magnetization and smoothed using a fourth-order Savitzky-Golay filter over 40 data points.

\section{Theoretical Framework}
\label{Theo}

For the description of the reorientation of the helix in cubic chiral magnets with weak anisotropy we consider an effective theory in the limit of small spin-orbit coupling $\lambda_{\mathrm{SOC}}$ for which $H_{c1}/H_{c2} \sim \lambda_{\mathrm{SOC}}^{2} \ll 1$. In this limit, the helix orientation can be conveniently described in terms of a Landau potential $\mathcal{V}$ for the helix vector $\vec{Q}$ only.

The properties of the magnetization, which are determined by the full Ginzburg-Landau functional~\cite{1980:Bak:JPhysCSolidState, 1980:Nakanishi:SolidStateCommun}, mainly enter via the magnetic susceptibility tensor $\chi_{ij}$ that influences the stiffness of the helix orientation $\vec{Q}$ at a finite field $\vec{H}$. Importantly, this susceptibility is dominated by the magnetization of the pristine helix, while slight deformations of the helix magnetization due to crystalline anisotropies only lead to small corrections that are suppressed by powers of $\lambda_{\mathrm{SOC}}$. These corrections have been observed experimentally for particular field configurations in terms of higher harmonics, $\mathrm{e}^{\pm\mathrm{i}2\vec{Q}\vec{r}}$, with very small amplitude~\cite{1989:Lebech:JPhysCondensMatter, 2006:Grigoriev:PhysRevB2, 2014:Kousaka:JPSConfProc}. As a consequence, for small $\lambda_{\mathrm{SOC}}$, the helix magnetization basically remains undeformed during the reorientation process. 

In the following we present the general form of the effective Landau potential for the helix vector in the limit of small magneto-crystalline anisotropies in Sec.~\ref{Landau}. The latter are constrained by the symmetries of the crystal structure. We describe how to infer the parameters of the potential from experiment, notably the transverse and longitudinal susceptibilities, giving specific values for the case of MnSi. We then derive the expected trajectories of the helix pitch vector as a function of the magnetic field strength for different field directions in Sec.~\ref{Traject}. These considerations allow the identification of the residual Ising symmetries for specific field directions. In doing so we distinguish in particular those situations when starting from a single-domain and a multi-domain state, corresponding to experiments after (high-)field cooling and zero-field cooling, respectively.

\subsection{Effective Landau potential for the helix axis}
\label{Landau}

The magnetization of a helix $\vec{M}(\vec{r}) = M_{s}(\hat{e}_{1}\cos(\vec{Q}\vec{r}) + \hat{e}_{2}\sin(\vec{Q}\vec{r}))$ with amplitude $M_{s}$ is determined by the orthonormal basis $\hat{e}_{i}\hat{e}_{j} = \delta_{ij}$, that is, for instance, right-handed $\hat{e}_{1} \times \hat{e}_{2} = \hat{e}_{3} \equiv \vec{Q}/|\vec{Q}|$ for a right-handed helix. The helix is defined up to a $U(1)$ phase corresponding to rotations of $\hat{e}_{1}$ and $\hat{e}_{2}$ around the $\hat{e}_{3}$ axis. Moreover, the helix is invariant under the transformation $\vec{Q} \to -\vec{Q}$ and $\hat{e}_{2} \to -\hat{e}_{2}$, and, in this sense, $\vec{Q}$ can be effectively considered as a director. The size of $\vec{Q}$ is determined by the Dzyaloshinskii-Moriya interaction that is proportional to spin-orbit coupling $\lambda_{\mathrm{SOC}}$ and weak in MnSi. For small $\lambda_{\mathrm{SOC}}$, we can expand the Landau potential $\mathcal{V}$ in a Taylor series in $\vec{Q} \propto \lambda_{\mathrm{SOC}}$ and confine ourselves to the lowest order terms only. Moreover, as the amplitude of $\vec{Q}$ is basically fixed we concentrate on the orientation $\hat{Q} = \vec{Q}/|\vec{Q}|$ that we treat as a director so that the potential $\mathcal{V}(\hat{Q})$ should be an even function of $\hat{Q}$.

The potential at zero field is attributed to the magneto-crystalline anisotropies. The latter are often referred to as cubic anisotropies in MnSi, but they are governed in fact by the tetrahedral point group $T$ of its cubic space group $P2_{1}3$. The tetrahedral symmetries contain a two-fold rotation symmetry $C_{2}$ around a cubic axis $\langle100\rangle$ and a three-fold rotation symmetry $C_{3}$ around $\langle111\rangle$~\footnote{Actually, in point group $T$ the direction $[111]$ is crystallographically equivalent only to $[1\bar{1}\bar{1}]$, $[\bar{1}1\bar{1}]$, and $[\bar{1}\bar{1}1]$. The directions $[\bar{1}\bar{1}\bar{1}]$, $[\bar{1}11]$, $[1\bar{1}1]$, and $[11\bar{1}]$ are also equivalent, but distinct from $[111]$. As the helix orientation in MnSi, however, is determined by a director rather than by a vector, i.e., $+\vec{Q}$ and $-\vec{Q}$ may not be distinguished, in the following the notation {$\langle111\rangle$} refers to both classes of crystallographic directions.}. The corresponding potential for $\hat{Q}$ consistent with these symmetries reads
\begin{align} \label{TPot}
\mathcal{V}_{T}(\hat{Q}) &= \varepsilon^{(1)}_{T} (\hat{Q}_{x}^{4} + \hat{Q}_{y}^{4} + \hat{Q}_{z}^{4}) \\ \nonumber
 &+ \varepsilon^{(2)}_{T} (\hat{Q}_{x}^{2} \hat{Q}_{y}^{4} + \hat{Q}_{y}^{2} \hat{Q}_{z}^{4} + \hat{Q}_{z}^{2} \hat{Q}_{x}^{4}) + ...~.
\end{align}
The leading first term with energy density $\varepsilon^{(1)}_{T} \sim \lambda_{\mathrm{SOC}}^{4}$ is fourth order in spin-orbit coupling. Note that the other quartic invariant $(\hat{Q}_{x}^{2} \hat{Q}_{y}^{2} + \mathrm{cycl.})$ is redundant since it is up to a constant equivalent to the first term. Importantly, the term $\varepsilon^{(1)}_{T}$ is still invariant under a four-fold rotation, $C_{4}$, around one of the cubic axes and this symmetry is not contained in $T$. The emergent symmetry of the potential present in leading order in $\lambda_{\mathrm{SOC}}$ is broken in the next-to-leading order by the second term with $\varepsilon^{(2)}_{T} \sim \lambda_{\mathrm{SOC}}^{6}$. The other terms of order $\mathcal{O}(\lambda_{\mathrm{SOC}}^{6})$, i.e., $\hat{Q}_{x}^{2} \hat{Q}_{y}^{2} \hat{Q}_{z}^{2}$ and $(\hat{Q}_{x}^{6} + \mathrm{cycl.})$, preserve the $C_{4}$ symmetry and are less important. These invariants as well as terms of higher order are represented by the dots in Eq.~\eqref{TPot} and will be neglected in the following. 

In small fields, the Zeeman energy $\mathcal{V}_{H}(\hat{Q}) = -\frac{\mu_{0}}{2}\chi_{ij}\vec{H}_{i}\vec{H}_{j}$ is determined by the susceptibility tensor $\chi_{ij}$ of the helix magnetization for a \textit{fixed} pitch vector $\vec{Q}$. It reads explicitly
\begin{align} \label{HPot}
\mathcal{V}_{H}(\hat{Q}) = -\frac{\mu_{0}}{2} \left(\chi_{\perp} \vec{H}^{2} + (\chi_{\parallel}-\chi_{\perp})(\vec{H}\hat{Q})^{2} + ... \right).
\end{align}
In leading order in $\lambda_{\mathrm{SOC}}$, the Zeeman energy is governed by the susceptibility tensor of the pristine helix, $\chi_{ij} = \chi_{\parallel}\hat{Q}_{i}\hat{Q}_{j} + \chi_{\perp}(\delta_{ij}-\hat{Q}_i\hat{Q}_j)$, which is characterized by the susceptibilities longitudinal and transversal to the pitch vector, $\chi_{\parallel}$ and $\chi_{\perp}$, respectively. For a spherical sample with demagnetization factor $N = 1/3$, they are given by $\chi_{\nu} = \chi^{\mathrm{int}}_{\nu}/(1 + \chi^{\mathrm{int}}_{\nu}/3)$ with $\nu = \parallel,\perp$, where the internal susceptibilities were evaluated in Ref.~\onlinecite{2013:Janoschek:PhysRevB}. Deep inside the helimagnetically ordered phase at $\vec{H} = 0$, they obey $\chi^{\mathrm{int}}_{\parallel} = 2\chi^{\mathrm{int}}_{\perp}$ with $\chi^{\mathrm{int}}_{\parallel} \approx 0.34$ for MnSi~\cite{2015:Schwarze:NatureMater} resulting in the numerical values
\begin{align} \label{Suscept}
\chi_{\parallel} \approx 0.31,\quad \chi_{\perp} \approx 0.16~.
\end{align}
The first term on the right-hand side of Eq.~\eqref{HPot} is independent of $\hat{Q}$ but is kept here as it allows us to compute quantitatively the field dependence of the homogeneous magnetization across the transition. 

The Zeeman potential in Eq.~\eqref{HPot} at this order is still invariant with respect to an arbitrary simultaneous rotation of both $\vec{H}$ and $\hat{Q}$. This symmetry will be broken by shape anisotropies due to demagnetization fields in non-spherical samples; a situation we do not consider here. More interestingly, crystalline anisotropies also reduce this rotation symmetry and modify the susceptibility giving rise, for example, to an additional term $(H_{x}^{2} \hat{Q}_{y}^{2} + \mathrm{cycl.})$ in the Zeeman potential. An analysis of the Ginzburg-Landau theory~\cite{1980:Bak:JPhysCSolidState, 1980:Nakanishi:SolidStateCommun} for the reorientation transition at $H_{c1}$ shows that this term is of order $\mathcal{O}(\lambda^{6}_{\mathrm{SOC}})$. It is thus similarly important for the description of the helix reorientation as $\varepsilon^{(2)}_{T}$ of Eq.~\eqref{TPot}. However, within the accuracy of our experiments, we were not able to distinguish unambiguously between the various terms of order $\mathcal{O}(\lambda^{6}_{\mathrm{SOC}})$ so that we will neglect such corrections to Eq.~\eqref{HPot} for simplicity.

The total mean-field potential for the pitch orientation $\mathcal{V} = \mathcal{V}_{T} + \mathcal{V}_{H}$ depends on the parameters $\varepsilon^{(1)}_{T}$ and $\varepsilon^{(2)}_{T}$. From an analysis of our SANS data in MnSi at $T = 5$~K we find that the values 
\begin{align} \label{Epsilons}
\varepsilon_{T}^{(1)} \approx 0.0034~\mu{\mathrm{eV}/\textrm{\AA}}^{3},\quad \varepsilon_{T}^{(2)} \approx 0.35\,\varepsilon^{(1)}_{T}
\end{align}
provide an excellent description of the data. These values will be used in the following discussion.

We note that Plumer and Walker \cite{1981:Plumer:JPhysCSolidStatePhys} as well as Kataoka and Nakanishi \cite{1981:Kataoka:JPhysSocJpn} already developed an account for the helix orientation in MnSi on the level of the magnetization. These authors, however, neglected contributions to the free energy density akin to $\varepsilon^{(2)}_{T}$ in Eq.~\eqref{TPot} breaking the $C_{4}$ rotation symmetry and incorrectly computed the transverse susceptibility $\chi_{\perp}$ entering Eq.~\eqref{HPot}. Walker \cite{1989:Walker:PhysRevB} subsequently predicted the two phase transition for a field along $\langle 100\rangle$ but only within a stability analysis around this field orientation. Taken together with the additional aspects covered in the present study, our description of the helix orientation in terms of $\hat{Q}$ goes well beyond the work reported in Refs.~\onlinecite{1981:Plumer:JPhysCSolidStatePhys, 1981:Kataoka:JPhysSocJpn, 1989:Walker:PhysRevB}.

\subsection{Helix trajectories and elastic Ising transitions}
\label{Traject}

\begin{figure}
\includegraphics[width=1.0\linewidth]{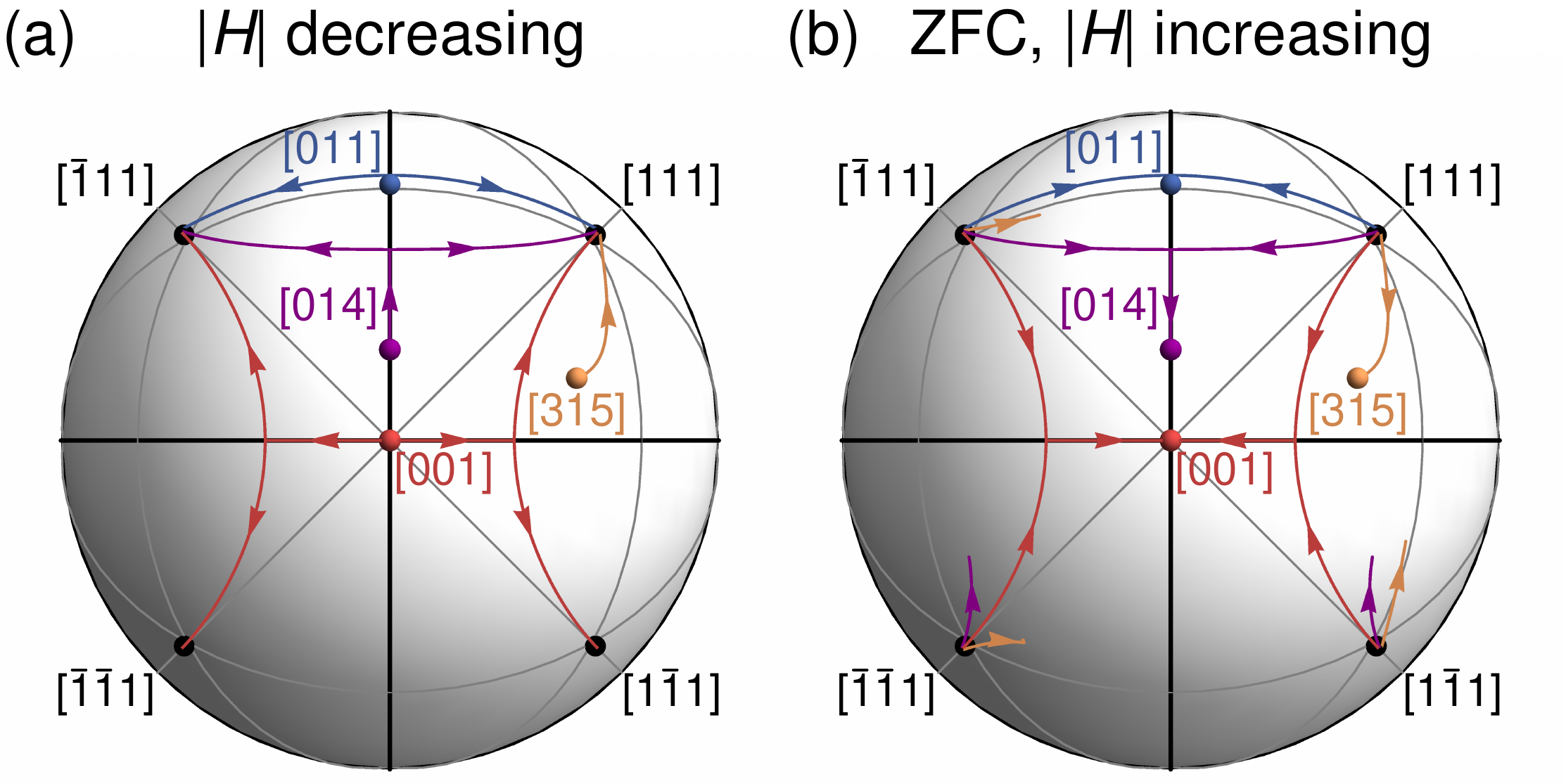}
\caption{Trajectories of the pitch vector $\hat{Q}$ on the unit sphere for $\varepsilon^{(2)}_{T}/\varepsilon^{(1)}_{T} = 0.35$ and four different field directions $\vec{H}$ (colored dots). (a)~Trajectories for decreasing fields starting from high fields with $\hat{Q}\parallel\vec{H}$. Bifurcations indicate elastic Ising transitions with the accompanying phase separation. (b)~Trajectories for increasing fields starting from the zero-field cooled state with all $\langle111\rangle$ domains (black dots) being equally populated. Trajectories starting from helimagnetic domains, whose $\hat{Q}$ enclose larger angles with $\hat{H}$, are discontinuous characteristic of first-order transitions.}
\label{figure2}
\end{figure}

The pitch orientation for an applied magnetic field $\vec{H}$ may be determined by minimizing the total Landau potential $\mathcal{V} = \mathcal{V}_{T} + \mathcal{V}_{H}$ with respect to $\hat{Q}$. In zero magnetic field, the pitch orientation is determined by the magneto-crystalline anisotropies $\mathcal{V}_{T}$ of Eq.~\eqref{TPot}. For MnSi, the prefactor of the first term is positive, $\varepsilon^{(1)}_{T} > 0$, such that it is minimized for $\hat{Q} \parallel \langle111\rangle$. Thus, four domains are formed under zero-field cooling, each of which is defined up to an $U(1)$ phase as mentioned above. For large fields, on the other hand, the pitch orientation will be aligned along $\vec{H}$ to minimize the Zeeman energy $\mathcal{V}_{H}$. When decreasing the magnitude of the field, the pitch vector $\hat{Q}$ will reorient towards one of the $\langle111\rangle$ directions. For generic field orientations, where $\hat{H}$ belongs to one of the octants of the unit sphere centered around $\langle111\rangle$ and separated by the black great circles in Fig.~\ref{figure1}(b), the pitch vector smoothly reorients towards the corresponding $\langle111\rangle$ direction, which is illustrated in Fig.~\ref{figure2}(a) for $\vec{H}\parallel[315]$.

An exception of this generic behavior is observed for field directions along the great circles connecting the $\langle100\rangle$ axes shown in black. In these cases, the four $\langle111\rangle$ domains may be grouped into two energetically degenerate pairs whose $\vec{Q}$ encloses the same angle with the field direction. Consequently, a continuous (second-order) phase transition is expected as a function of field strength at a well-defined critical field $H_{c1}$. In order to determine the detailed characteristics of this transition, we perform a stability analysis of $\hat{Q}$ around the field direction, e.g., $\hat{H} = (0,\sin\alpha,\cos\alpha)$ parametrized by the polar angle $\alpha$. We set $\hat{Q} = \hat{H} \sqrt{1 - x_{1}^{2} - x_{2}^{2}} + x_{1} \hat{v}_{1}+x_{2} \hat{v}_{2}$ with the orthonormal vectors $\hat{v}_{1} = (1,0,0)$ and $\hat{v}_{2} = (0,\cos\alpha,-\sin\alpha)$. Thus, the coordinates $x_{1}$ and $x_{2}$ describe the deviation away and along the great circle, respectively. Expanding the potential $\mathcal{V}$ in $x_{1}$ and $x_{2}$ allows to readily identify an Ising instability. 

The direction of the Ising instability is along the direction of $x_{1}$ and away from the great circle $(0,\sin\alpha,\cos\alpha)$, provided that $\vec{H}$ does not point along a cubic $\langle100\rangle$ axis. For decreasing field magnitude, at $H_{c1}$ the pitch vector $\hat{Q}$ has to decide along which of the two directions away from the great circle it moves, i.e., whether $x_{1} > 0$ or $x_{1} < 0$, stabilizing a helimagnetic domain either along $[111]$ or $[\bar{1}11]$. This finding identifies the coordinate $x_{1}$ as an Ising order parameter of the transition. As typical examples we show trajectories of $\hat{Q}$ for $\vec{H}\parallel[011]$ and $[014]$ in Fig.~\ref{figure2}(a). Both trajectories bifurcate at the critical field $H_{c1}$ indicating possible phase separation into two domains.

A special situation arises for $\vec{H}\parallel\langle100\rangle$, where the field direction coincides with the crossing point of the two great circles shown in black in Fig.~\ref{figure1}(b). To lowest order in spin-orbit coupling $\lambda_{\mathrm{SOC}}$, i.e., for $\varepsilon_{T}^{(2)} = 0$, the potential for $\hat{Q}$ still possesses the $C_{4}$ symmetry that is reflected in a $\mathds{Z}_{4}$ symmetry of the effective theory for the vector $(x_{1}, x_{2})$. At this order, one would expect the transition to be described by a four-state clock model. However, the presence of $\varepsilon_{T}^{(2)}$  lowers the symmetry down to $\mathds{Z}_{2} \times \mathds{Z}_{2}$ and favors either the $x_{1}$ or $x_{2}$ direction depending on the sign of $\varepsilon_{T}^{(2)}$. As a consequence, the single $\mathds{Z}_{4}$ transition for $\varepsilon^{(2)}_{T} = 0$ splits into two subsequent $\mathds{Z}_{2}$ Ising transitions, in agreement with Ref.~\onlinecite{1989:Walker:PhysRevB}. This finding is illustrated by the trajectory shown in red in Fig.~\ref{figure2}(a). For decreasing field, a first instability is reached at $H^{[001]}_{c1,>}$. Here, the pitch vector tilts along one of the $x_{1}$ directions for $\varepsilon^{(2)}_{T} > 0$. When the field is reduced further, $\hat{Q}$ tilts along one of the $x_{2}$ directions at a second instability at $H^{[001]}_{c1,<}$.

Whereas the trajectories are always continuous when decreasing the field, a different situation arises after zero-field cooling when all $\langle111\rangle$ domains are equally populated. In this case, the helix reorientation as a function of \textit{increasing} field is partially discontinuous. As illustrated in Fig.~\ref{figure2}(b), only the domains closest to the applied field direction $\hat{H}$ reorient smoothly, while the trajectories starting from the other domains are discontinuous. The latter trajectories are shown up to their spinodal point where they terminate, signaling a jump into the stable domain configuration. 

\begin{figure}
\includegraphics[width=1.0\linewidth]{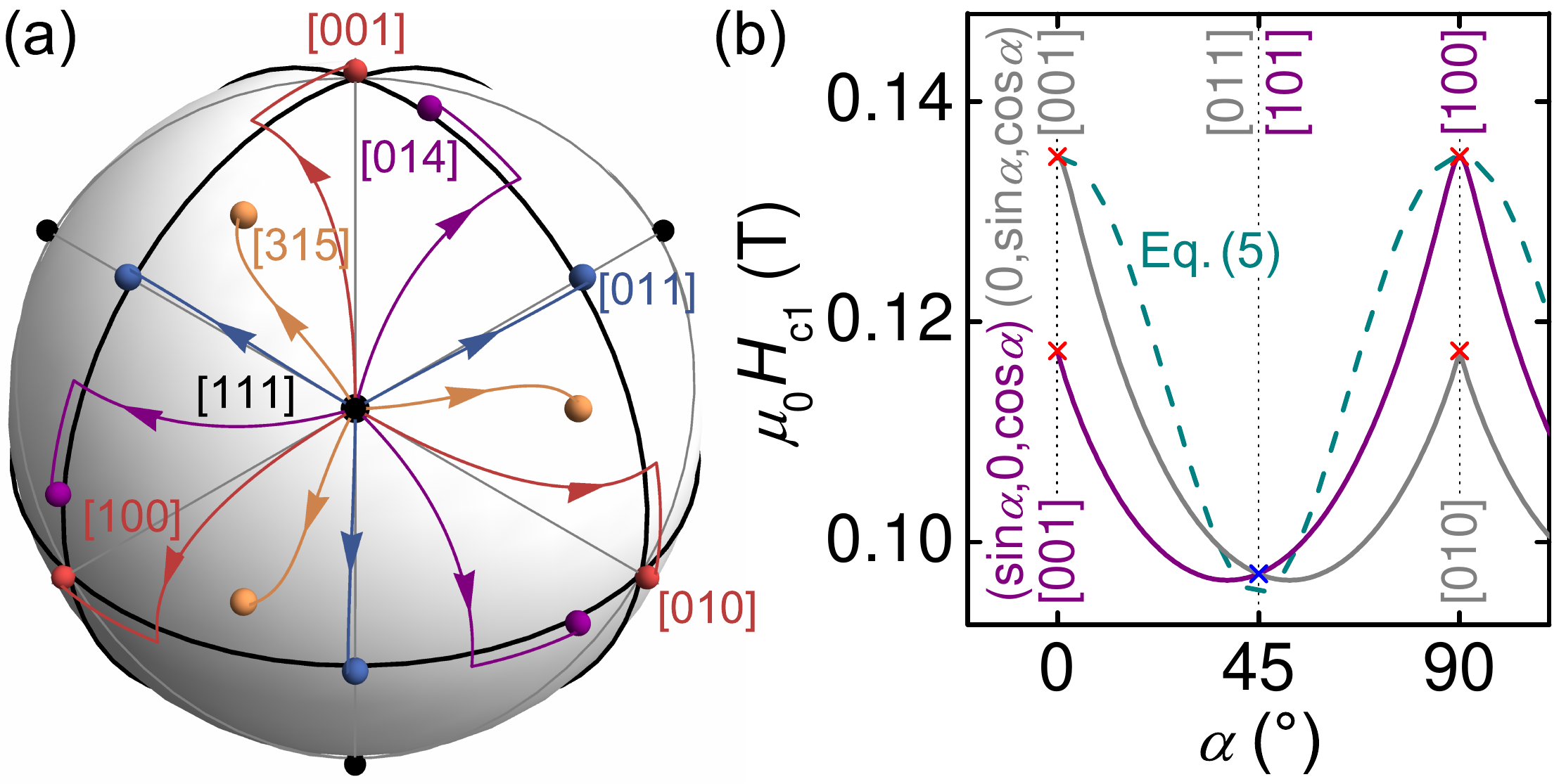}
\caption{Helix trajectories and critical fields. (a)~Trajectories of the pitch vector $\hat{Q}$ starting from the [111] domain in zero field (central black dot) for increasing fields along different directions $\vec{H}$ (colored dots). Trajectories of the same color are related by the $2\pi/3$ rotations around $[111]$ in accordance with the point group $T$. (b)~Critical field $H_{c1}$ as a function of the angle $\alpha$ for $\hat{H} = (0,\sin\alpha,\cos\alpha)$ and $\hat{H} = (\sin\alpha,0,\cos\alpha)$. The limiting values for $\alpha \to 0$ indicate the critical fields of the two elastic Ising transitions, $H^{[001]}_{c1,>}$ and $H^{[001]}_{c1,<}$. The dashed line corresponds to the situation for $\varepsilon^{(2)}_{T} = 0$.}
\label{figure3}
\end{figure}

The low symmetry of the tetrahedral point group $T$ is further illustrated when considering the evolution of the $[111]$ domain again for increasing field $\vec{H}$ pointing along $[hkl]$, $[lhk]$, and $[klh]$ ($k,l,h > 0$), see Fig.~\ref{figure3}(a). The resulting trajectories of $\hat{Q}$ are related by a $2\pi/3$ rotation symmetry around $[111]$. Correspondingly, fields along cyclically permuted directions $[hk0]$, $[0hk]$, and $[k0h]$ yield identical transition fields $H_{c1}$, while being different from the values for $\hat{H}$ along $[kh0]$, $[0kh]$, and $[h0k]$. 

This difference is illustrated in Fig.~\ref{figure3}(b) where we compare the evolution of $H_{c1}$ for fields $\vec{H}$ applied along $(0,\sin\alpha,\cos\alpha)$ and $(\sin\alpha,0,\cos\alpha)$ as a function of the angle $\alpha$. For general values of $\alpha$, the critical field values differ, with exception of $\vec{H} \parallel \langle110\rangle$. For $\vec{H} \parallel \langle100\rangle$, a special situation arises as the limiting values of $H_{c1}(\alpha)$ for $\alpha \to 0$ identify the upper and lower critical field of the $\mathds{Z}_{2} \times \mathds{Z}_{2}$ transition, $H^{[001]}_{c1,>}$ and $H^{[001]}_{c1,<}$, respectively. For comparison, the dashed line in Fig.~\ref{figure3}(b) shows the critical field obtained for $\varepsilon_{T}^{(2)} = 0$, which reads
\begin{align} \label{Hc1}
H_{c1}(\alpha)|_{\varepsilon^{(2)}_{T} = 0} = \sqrt{\frac{\varepsilon^{(1)}_{T}(3 + \cos(4\alpha))}{\mu_{0}(\chi_{\parallel}-\chi_{\perp})}}~.
\end{align}
In this approximation, the critical fields along $\langle100\rangle$ and $\langle110\rangle$ satisfy the ratio $H^{\langle100\rangle}_{c1} \approx \sqrt{2} H^{\langle110\rangle}_{c1}$, as previously pointed out in Ref.~\onlinecite{1981:Plumer:JPhysCSolidStatePhys}.

Note that the transitions are described by an Ising order parameter but do not belong to the three-dimensional Ising universality class. The reorientation transition of the helix is an elastic transition that is quite distinct from conventional phase transitions in magnets. Already a slight reorientation of the helix involves a macroscopic reconstruction of the magnetization. The rotation of the pitch vector $\vec{Q}$ by a small angle $\delta$ within a macroscopic domain of linear size $L$ requires large, non-perturbative changes of the magnetization over distances of order $L\delta  \gg 1$ at the domain boundary. A similar situation arises at continuous symmetry-breaking elastic transitions of atomic crystals~\cite{2015:Zacharias:EurPhysJSpecialTopics, 2015:Zacharias:PhysRevLett}. In the latter systems, the phonons soften at the transition but only along a particular direction in momentum space. A preliminary analysis indicates that the low-energy excitations of the helix, the helimagnons~\cite{2010:Janoschek:PhysRevB, 2015:Kugler:PhysRevLett}, soften similarly only within a reduced subspace at the helix reorientation transitions.


\section{Experimental Results}
\label{ExpRes}

Experimentally, we have addressed the helix reorientation in MnSi by means of small-angle neutron scattering  as well as magnetisation and ac susceptibility measurements, where we focused on temperatures well below to onset of helimagnetic order. We begin our presentation with neutron scattering for magnetic fields applied along the high-symmetry directions $\langle110\rangle$ and $\langle100\rangle$. Rotating the sample with respect to the neutron beam, allows us to map the relevant part of reciprocal space in three dimensions and to track the trajectories of the pitch vector as a function of field. We find bifurcations as well as distinct differences between zero-field cooling and high-field cooling. Comparing our data with the model described in the previous subsection, we are able to quantitatively determine the strength of the anisotropy factors $\varepsilon^{(1)}_{T}$ and $\varepsilon^{(2)}_{T}$. In addition, discrepancies between the behavior for decreasing and increasing field magnitudes suggest an important influence of disclinations, as will be discussed in detail in Sec.~\ref{Discuss}.

In the second part of this section, we present the susceptibility as calculated from the measured magnetisation as well as measured directly, notably the ac susceptibility. We find that these data to be perfectly consistent with both our model and the neutron scattering results. By means of our theoretical description, we are able to explain quantitatively the different signatures observed in the susceptibility calculated from the measured magnetization and the measured ac susceptibility in terms of the slow response of the helix vector $\vec{Q}$ to changes in the applied magnetic field. Data shown for a large number of field directions underscore the remarkable agreement between experiment and theory.

\subsection{Small-angle neutron scattering}
\label{SANS}

\begin{figure}
\includegraphics[width=1.0\linewidth]{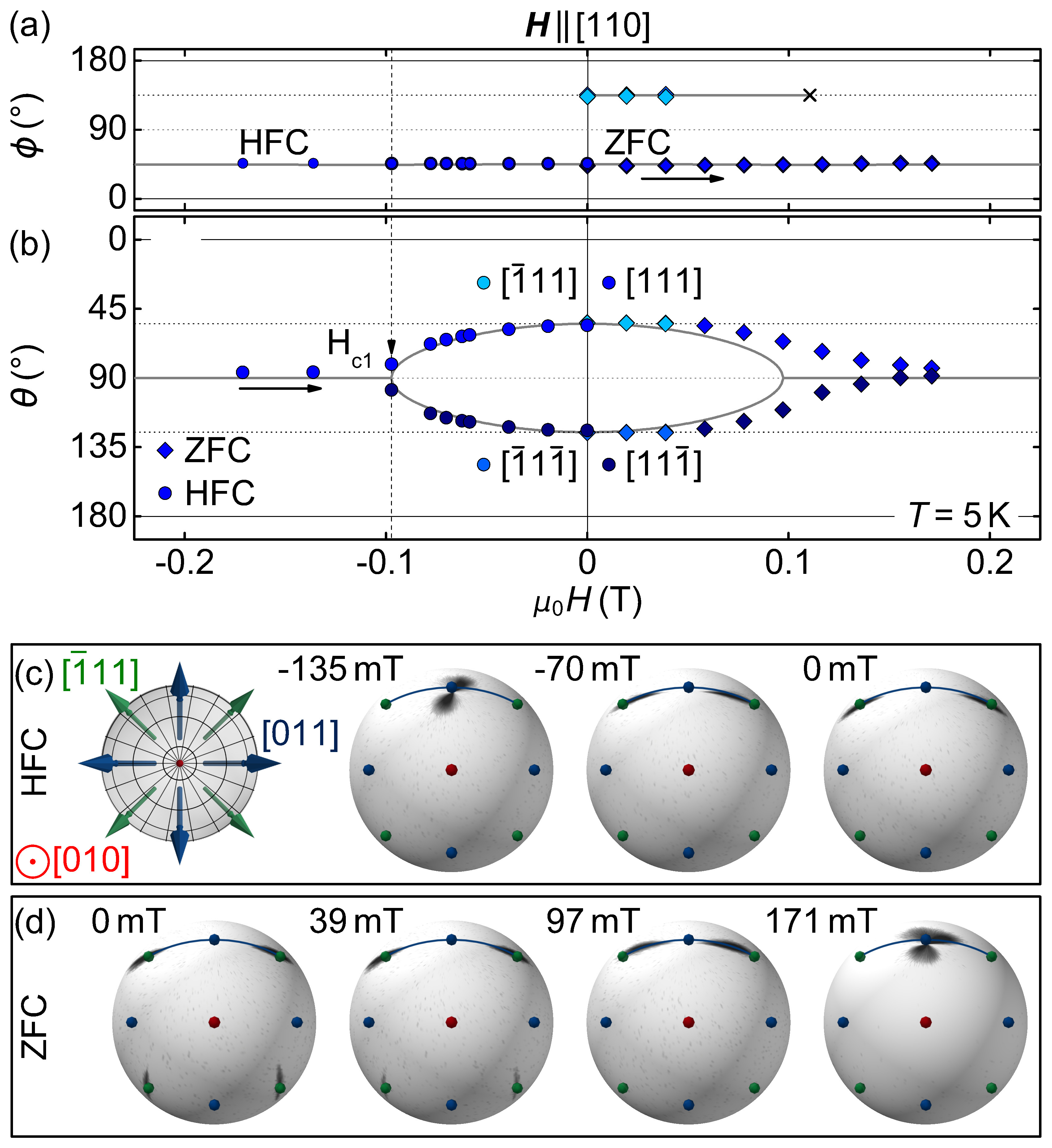}
\caption{Small-angle neutron scattering for $\vec{H}\parallel[110]$. \mbox{(a),(b)}~Position of intensity maxima for increasing field values (black arrows) shown on the unit sphere $\hat{Q} = (\sin\theta \cos\phi, \sin\theta \sin\phi, \cos\theta)$ parametrized by the angles $\phi$ and $\theta$. Starting from a single-domain conical state after high-field cooling (HFC), the trajectories bifurcate with a sharp critical signature at $\mu_{0}|H^{[110]}_{c1}| \approx 95$~mT (vertical dashed line), and multiple domains form due to phase separation. These signatures are smoothed when multiple domains merge, e.g., after zero-field cooling (ZFC). Solid gray lines are a fit to theory. (c)~Typical intensity distributions after HFC in large negative fields. (d)~Typical intensity distributions after ZFC. The colored points mark the high-symmetry directions $\langle111\rangle$ (green), $\langle110\rangle$ (blue), and $\langle100\rangle$ (red).}
\label{figure4}
\end{figure}

\begin{figure}
\includegraphics[width=1.0\linewidth]{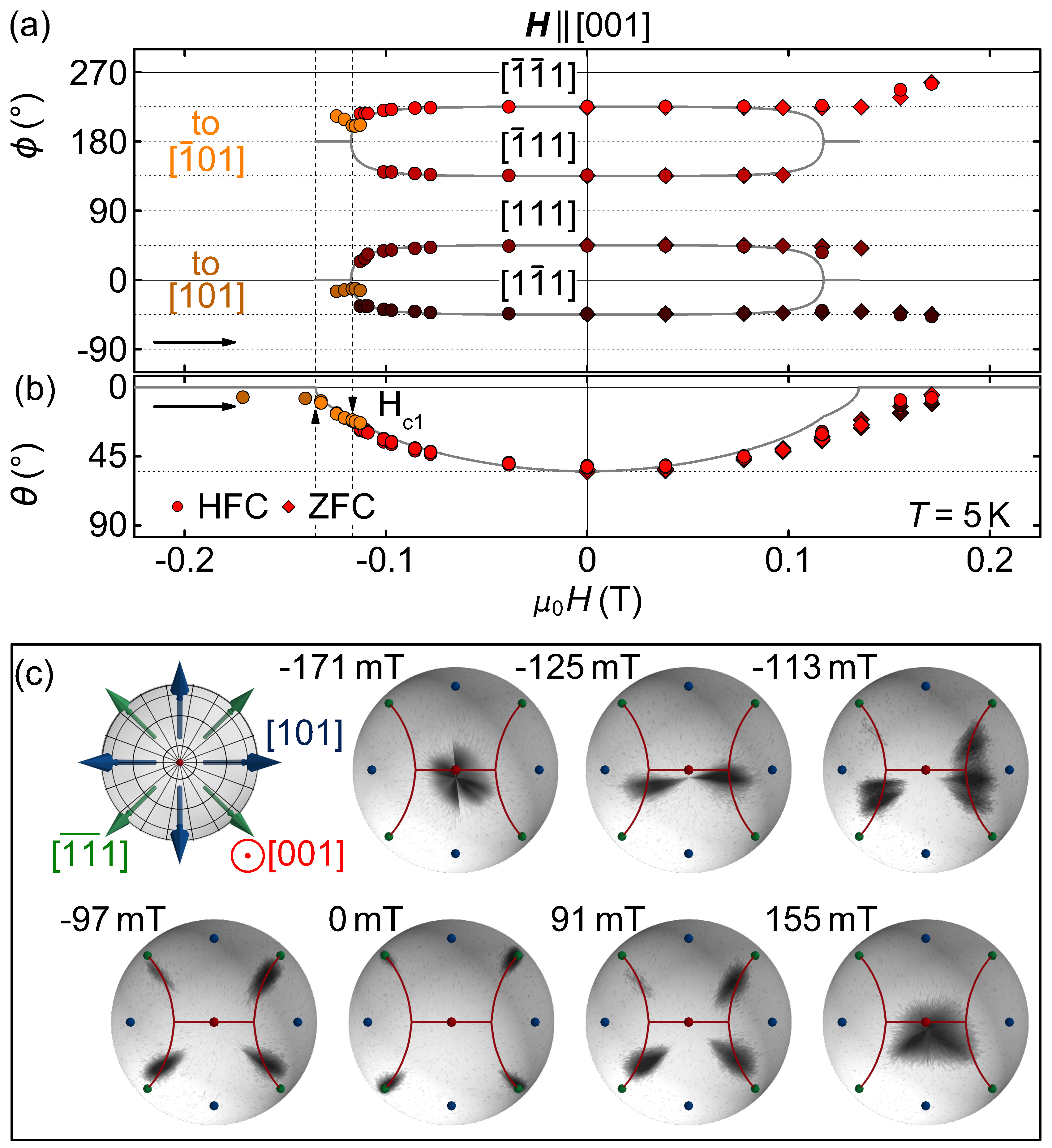}
\caption{Small-angle neutron scattering for $\vec{H}\parallel[001]$ in analogy to Fig.~\ref{figure4}. \mbox{(a),(b)}~Position of intensity maxima for increasing field values. Starting from a single-domain conical state after high-field cooling (HFC), two bifurcations are observed at the critical fields $\mu_{0}|H^{[001]}_{c1,>}| \approx 135$~mT and $\mu_{0}|H^{[001]}_{c1,<}| \approx 118$~mT (vertical dashed lines). No sharp signatures are observed when multiple domains merge, e.g., after zero-field cooling (ZFC). Solid gray lines are a fit to theory. (c)~Typical intensity distributions after HFC in large negative fields. The colored points mark the high-symmetry directions $\langle111\rangle$ (green), $\langle110\rangle$ (blue), and $\langle100\rangle$ (red).}
\label{figure5}
\end{figure}

By means of small-angle neutron scattering, we tracked the positions of the intensity maxima associated with the helimagnetic order ($|\vec{Q}| \approx 0.035~\textrm{\AA}^{-1}$) across the helix reorientation as a function of field. At a given temperature and magnetic field value, a series of two-dimensional scattering patterns was recorded while the sample was rotated by $180^{\circ}$ in $1^{\circ}$ steps. From these patterns we constructed the three-dimensional intensity distributions depicted in Figs.~\ref{figure4} and \ref{figure5}. As intensity at $\hat{Q}$ and $-\hat{Q}$ arises from the same helical domain, we averaged over both maxima and analyzed the behavior on one hemisphere. For further information on our analysis and the consequences of experimental misalignment, we refer to the supplemental material~\cite{supplement}.

In our SANS measurements we focused on two field configurations, namely $\vec{H}\parallel[110]$ and $\vec{H}\parallel[001]$, both of which are expected to show elastic Ising transitions. Measurements were performed starting either from a high-field cooled (HFC) single-domain conical state with $\hat{Q}\parallel\vec{H}$ at high negative fields or from a zero-field cooled (ZFC) state with equally populated helical domains with $\hat{Q}$ pointing along one of the $\langle111\rangle$ axes. In both cases, data were recorded for increasing field values. In the following, we use the parametrization $\hat{Q} = (\sin\theta \cos\phi, \sin\theta \sin\phi, \cos\theta)$ in order to describe the position of the intensity maxima as observed on a unit sphere.

We begin with the properties for magnetic field parallel to $[110]$ shown in Fig.~\ref{figure4}. In a large negative field, the conical state is observed with $\hat{Q}\parallel\vec{H}$ translating to $\theta = 90^{\circ}$ and $\phi = 45^{\circ}$. Upon lowering the field, nothing happens until the Ising transition may be identified by a sharp bifurcation in $\theta$ at negative $\mu_{0} H^{[110]}_{c1} \approx 95$~mT. The bifurcation indicates a phase separation into two helical domains approaching $[111]$ ($\theta = 54.7^{\circ}$) or $[11\bar{1}]$ ($\theta = 125.3^{\circ}$), respectively. The gray solid line represents a fit considering Eqs.~\eqref{TPot} and \eqref{HPot}. It is in excellent agreement with the data. Note that the theoretical curve for $\phi$ slightly deviates from $45^{\circ}$, less than $0.5^{\circ}$ close to $H_{c1}$, due to the finite $\varepsilon^{(2)}_{T}$.

In contrast, increasing the field to positive values starting from the multi-domain state at $H = 0$, the sharp critical signatures are smeared and the trajectories substantially deviate from the theoretical prediction. After zero-field cooling, at $H = 0$  the domains at $[\bar{1}1\bar{1}]$ ($\theta = 125.3^{\circ}$, $\phi = 135^{\circ}$) and $[\bar{1}11]$ ($\theta = 54.7^{\circ}$, $\phi = 135^{\circ}$) are also populated. For $H > 0$ these domains become metastable with the corresponding intensity vanishing around ${\sim}50$~mT indicating first-order transitions. The value of this depopulation field is about half the value of the spinodal point predicted theoretically as marked by the cross in Fig.~\ref{figure4}(a).  

We turn now to the situation for a magnetic field along $[001]$ shown in Fig.~\ref{figure5}. In a large negative field, again the conical state is observed with $\hat{Q}\parallel\vec{H}$ translating to $\theta = 0^{\circ}$ and an arbitrary value of $\phi$. Upon lowering the field, the first Ising transition is observed at negative $\mu_{0}H_{c1,>}^{[001]} \approx 135$~mT, where $\theta$ becomes finite and two values of $\phi$ may be defined, namely $\phi = 0^{\circ}$ and $\phi = 180^{\circ}$. A further bifurcation of $\phi$ marks the second Ising transition at $\mu_{0}H^{[001]}_{c1,<} \approx 118$~mT, accompanied by a small kink in $\theta$. As a result, at low fields all four helical domains are populated; $[111]$ ($\phi = 45^{\circ}$), $[\bar{1}11]$ ($\phi = 135^{\circ}$), $[\bar{1}\bar{1}1]$ ($\phi = 225^{\circ}$), and $[1\bar{1}1]$ ($\phi = 315^{\circ} \mathrel{\widehat{=}} -45^{\circ}$), all at $\theta = 54.7^{\circ}$.

When increasing the field to positive values starting from a multi-domain state at $H = 0$ results in qualitatively different behavior. Instead of two subsequent Ising transitions, all four domains smoothly reorient towards the field direction in a direct trajectory as indicated by a constant $\phi$ and a smooth decrease of $\theta$. After zero-field cooling, the same behavior is observed. Critical signatures are absent when the multiple domains merge to a single-domain conical state at high fields, similar to the situation for increasing field values along $[110]$ in Fig.~\ref{figure4}.

Closer inspection of the neutron scattering data reveals that the magnetic field in the experiment was misaligned by a few degrees, see supplement for a detailed discussion~\cite{supplement}. Due to the spherical coordinate system used, overall small misalignment angles around $\theta = 0$, however, may translate to putatively large deviations in $\phi$. This instance explains, in particular, the comparably large discrepancy between the first couple of experimental data points at large negative fields in Fig.~\ref{figure5}(a) and the calculated values of $\phi = 0^{\circ}$ and $\phi = 180^{\circ}$, respectively. In addition, it is noteworthy that we still observe phase separation at the transitions despite of certain domains being slightly favored due to the misalignment. The latter also smooths out the signatures of the reorientation and, in fact, theoretical calculations indicate quite substantial smearing for the misalignment angles of our experiment. The observation of relatively sharp features in our neutron scattering data is therefore quite unexpected. We return to this issue in further detail below.


\subsection{Magnetic susceptibility}
\label{MagSus}

\begin{figure}
\includegraphics[width=1.0\linewidth]{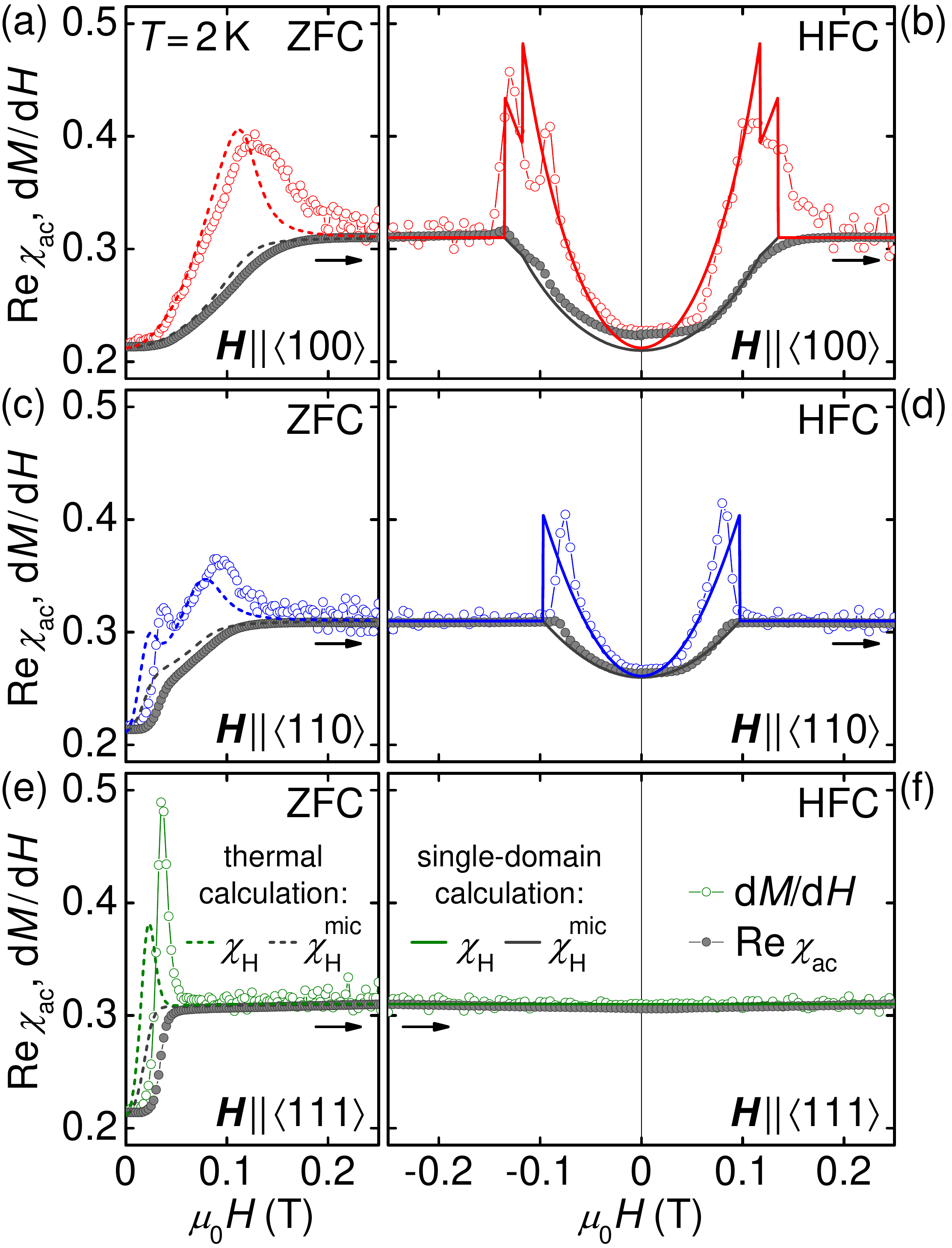}
\caption{Susceptibility as a function of increasing field. \mbox{(a),(b)}~Data for $\vec{H}\parallel\langle100\rangle$ after zero-field cooling (ZFC) and high-field cooling (HFC). We show the susceptibility calculated from the measured magnetization, $\mathrm{d}M/\mathrm{d}H$ (open symbols), and the real part of the ac susceptibility measured at 911~Hz, $\mathrm{Re}\,\chi_{\mathrm{ac}}$ (solid symbols). Dashed and solid lines are the results of our calculations, see text for details. \mbox{(c)--(f)}~Corresponding data for $\vec{H}\parallel\langle110\rangle$ and $\vec{H}\parallel\langle111\rangle$.}
\label{figure6}
\end{figure}

In the following we present measurements and calculations of the susceptibility around the helix reorientation complementing our neutron scattering results. We distinguish the susceptibility calculated from the measured magnetization, $\mathrm{d}M/\mathrm{d}H$, and the real part of the ac susceptibility, $\mathrm{Re}\,\chi_{\mathrm{ac}}$. Note that for MnSi the susceptibility of the single-domain conical state at higher fields is characterized by a plateau of constant absolute value $\chi_\parallel \approx 0.31$, see Eq.~\eqref{Suscept}, while the multi-domain helical state at zero field is governed by the average response of all domains when populated equally yielding the reduced value $\chi_\perp + \frac{1}{3}(\chi_\parallel - \chi_\perp) \approx 0.21$~\cite{2013:Janoschek:PhysRevB}. 

We begin our description with the behavior for fields along $\langle100\rangle$ after zero-field cooling, shown in Fig.~\ref{figure6}(a). A broad maximum in $\mathrm{d}M/\mathrm{d}H$ (open symbols) may be attributed to the smooth reorientation of the four helical $\langle111\rangle$ domains towards the field direction. The maximum is not tracked by $\mathrm{Re}\,\chi_{\mathrm{ac}}$ measured at finite frequency (solid symbols) indicating the importance of slow dynamics, cf.\ Refs.~\onlinecite{2012:Bauer:PhysRevB, 2014:Levatic:PhysRevB, 2016:Bauer:PhysRevB}. When increasing the field starting from the conical state at large negative values, cf. Fig.~\ref{figure6}(b), a distinct double peak is observed in $\mathrm{d}M/\mathrm{d}H$ as the characteristic of the two subsequent Ising transitions. Again, $\mathrm{Re}\,\chi_{\mathrm{ac}}$ does not track $\mathrm{d}M/\mathrm{d}H$ but still exhibits two distinct kinks at $H^{[001]}_{c1,>}$ and $H^{[001]}_{c1,<}$. For increasing positive field values, a smeared maximum without sharp critical signatures in $\mathrm{d}M/\mathrm{d}H$ resembles the situation after zero-field cooling. Similar to the neutron scattering data, we thus observe hysteretic behavior close to the critical fields also in the magnetic susceptibility.

For fields along $\langle110\rangle$ after zero-field cooling shown in Fig.~\ref{figure6}(c) a small and slightly broadened peak preempts the maximum in $\mathrm{d}M/\mathrm{d}H$. This hump indicates the depopulation of the two helical domains which are energetically unfavored at a similar field value $\sim 50$mT, as previously observed in neutron scattering, cf. Fig.~\ref{figure4}(a). In $\mathrm{Re}\,\chi_{\mathrm{ac}}$, only a kink is observed at the corresponding field value. When starting in the conical state at negative fields, see Fig.~\ref{figure6}(d), the sharp maximum in $\mathrm{d}M/\mathrm{d}H$ may be associated with the single Ising transition expected for this field configuration. It is helpful to note the comparatively small discrepancy between the signature of the phase transition at positive and negative fields.

For fields along $\langle111\rangle$, see Fig.~\ref{figure6}(e), a relatively sharp maximum observed after zero-field cooling may be attributed to the simultaneous depopulation of three helimagnetic domains in favor of the fourth, where the latter domain is characterized by $\vec{Q} \parallel \vec{H}$. As illustrated in Fig.~\ref{figure6}(f), once this configuration, i.e., the conical state, is stabilized, the susceptibility remains constant. In other words, due to the magneto-crystalline anisotropies and Zeeman energy being simultaneously minimized, $\hat{Q}$ remains unchanged with the susceptibility $\chi_{\parallel}$.

Next, we turn to the calculations of the susceptibility shown by the dashed and solid lines in Fig.~\ref{figure6}. For the solid lines (right column), we consider a macroscopic single domain with pitch vector $\hat{Q}_{\mathrm{min}}(\vec{H})$ that minimizes $\mathcal{V}$ for a given field $\vec{H}$. Here the susceptibility follows from the Landau potential $\chi_{H} \equiv \mathrm{d}M/\mathrm{d}H = -\partial^{2}_{H} \mathcal{V}(\hat{Q}_{\mathrm{min}})/\mu_{0}$, and may be decomposed into two contributions
\begin{align}
\chi_{H} &= \chi^{\mathrm{mic}}_{H} + \chi^{\mathrm{mac}}_{H}, \\\nonumber 
\chi^{\mathrm{mic}}_{H} &= \hat{H}_{i} \chi_{ij} \hat{H}_j|_{\mathrm{min}} = \chi_\perp + (\chi_{\parallel} - \chi_{\perp}) (\hat{H} \hat{Q}_{\mathrm{min}})^{2}, \\\nonumber
\chi^{\mathrm{mac}}_{H} &= -\frac{1}{\mu_{0}} \frac{\partial^{2} \mathcal{V}(\hat{Q}_{\mathrm{min}})}{\partial \hat{Q}^{i}_{\mathrm{min}} \partial \hat{Q}^{j}_{\mathrm{min}}} \frac{\partial \hat{Q}^{i}_{\mathrm{min}}}{\partial H} \frac{\partial \hat{Q}^{j}_{\mathrm{min}}}{\partial H}~,
\end{align}
where $H = |\vec{H}|$. The first term, $\chi^{\mathrm{mic}}_{H}$, derives from the response of the microscopic magnetization of the helix for a \textit{fixed} pitch vector $\hat{Q}_{\mathrm{min}}$. The second term, $\chi^{\mathrm{mac}}_{H}$, accounts for the field dependence of the pitch vector, $\hat{Q}_{\mathrm{min}}(\vec{H})$. This corresponds to the reorientation of helimagnetic domains on macroscopic scales.

The two contributions are associated with very different time scales. While the (local) magnetization responds to changes of the magnetic field much faster than the typical time scales accessible by susceptibility measurements of about 0.1~ms, the macroscopic reorientation process is very slow implying a large characteristic time scale $\tau_{\hat{Q}}$. For variations at frequencies $f_{\mathrm{ac}} \gg 1/\tau_{\hat{Q}}$, the pitch vector $\hat{Q}$ remains unchanged as it is not able to follow the oscillating field and the corresponding contribution $\chi^{\mathrm{mac}}_{H}$ is suppressed. Thus, the ac susceptibility measured with an excitation frequency $f_{\mathrm{ac}}$ is given by $\mathrm{Re}\,\chi_{\mathrm{ac}} = \chi^{\mathrm{mic}}_{H}$ (gray lines in Fig.~\ref{figure6}). In contrast, the susceptibility calculated from the measured magnetization represents the static limit ($f_{\mathrm{ac}} = 0$) probing both contributions, $\mathrm{d}M/\mathrm{d}H = \chi_{H} = \chi^{\mathrm{mic}}_{H} + \chi^{\mathrm{mac}}_{H}$ (colored lines in Fig.~\ref{figure6}). 

Using the values given in Eqs.~\eqref{Suscept} and \eqref{Epsilons} obtained from a fit to the neutron scattering data, our calculations are in excellent agreement with the experimental data in Fig.~\ref{figure6}, where we find $f_{\mathrm{ac}} = 911~\mathrm{Hz} \gg 1/\tau_{\hat{Q}}$. In fact, in Ref.~\onlinecite{2012:Bauer:PhysRevB} it was demonstrated that $\mathrm{Re}\,\chi_{\mathrm{ac}}$ and $\mathrm{d}M/\mathrm{d}H$ remain distinct down to low frequencies even at rather high temperatures of 27.5~K where $(T_{c}-T)/T_{c} \approx 5\%$, providing an estimate for the lower bound of $\tau_{\hat{Q}} \geq 1$~s.

Furthermore, as the Zeeman potential $\mathcal{V}_{H}$ is quadratic in $\vec{H}$, the derivative $\partial_{H} \hat{Q}^{i}_{\mathrm{min}}$ is linear in $H$ for $H \to 0$ and, as a consequence, the zero-field limit of the susceptibility obeys $\chi_{H}|_{H = 0} = \chi^{\mathrm{mic}}_{H}$. For a single domain $\hat{Q}_{\mathrm{min}}\parallel[111]$, the value of $\chi_{H}|_{H = 0}$ expected theoretically amounts to 0.21, 0.26, or $0.31$ when the field $\hat{H}$ is applied along $\langle100\rangle$, $\langle110\rangle$, or $\langle111\rangle$, respectively. The experimental data are consistent with these values, where small deviations for $\vec{H}\parallel \langle 100\rangle$ are attributed to the misalignment of the field direction with respect to the crystalline $\langle100\rangle$ axes.

In contrast, after zero-field cooling all four helimagnetic domains will be populated with equal probability. With increasing field, in general, some of these domains become metastable and are expected to jump into the favored directions at first-order transitions. This process is observed at depopulation fields of ${\sim}50$~mT in neutron scattering, see Fig.~\ref{figure4}(a), as well as in the susceptibility inferred from the magnetization and the ac susceptibility, see left column of Fig.~\ref{figure6}. The description in terms of a single domain does not capture the behavior after zero-field cooling, see supplemental material for details~\cite{supplement}. Instead, we consider a simplistic model of thermally populated domains of a finite linear size $\xi_{\mathrm{dom}}$ with free energy density
\begin{align} \label{Thermal}
f = -\frac{k_{\mathrm{B}}T}{\xi_{\mathrm{dom}}^{3}} \log Z,\quad 
Z = \int \mathrm{d}\hat{Q}\,\mathrm{e}^{-\xi_{\mathrm{dom}}^{3}\mathcal{V}(\hat{Q})/(k_{\mathrm{B}}T)}~.
\end{align}
The resulting susceptibilities, $\langle\chi_{H}\rangle \equiv -\partial_{H}^{2} f/\mu_{0}$, are shown as colored dashed lines in the left column of Fig.~\ref{figure6}. A thermal energy density $k_{\mathrm{B}}T/\xi_{\mathrm{dom}}^{3} = 0.02\varepsilon^{(1)}_{T}$ was assumed corresponding to a linear length $\xi_{\mathrm{dom}} = 136~\textrm{\AA}$ at a temperature $T = 2$~K. The calculations qualitatively reproduce the additional signatures observed in $\mathrm{d}M/\mathrm{d}H$ for $\vec{H}\parallel\langle110\rangle$ and $\vec{H}\parallel\langle111\rangle$ close to the depopulation field. Note, however, that the estimate for $\xi_{\mathrm{dom}}$ is on the order of the helix wavelength in MnSi and hence unrealistically small. In fact, a more realistic model also should for instance, take into account the distribution of domain sizes and the influence of the domain walls.

At large excitation frequencies $f_{\mathrm{ac}}$, the susceptibility again is only sensitive to the response of the magnetization at fixed pitch vector. Accordingly, $\mathrm{Re}\,\chi_{\mathrm{ac}}$ may be described by a thermally averaged susceptibility $\langle\chi^{\mathrm{mic}}_{H}\rangle \equiv \chi_{\perp} + (\chi_{\parallel} - \chi_{\perp}) \langle(\hat{H}\hat{Q})^{2}\rangle$ (gray dashed lines), where $\langle\mathcal{O}\rangle = \int \mathrm{d}\hat{Q}\,\mathcal{O}\mathrm{e}^{-\xi_{\mathrm{dom}}^{3}\mathcal{V}(\hat{Q})/(k_{\mathrm{B}}T)} / Z$ and $k_{\mathrm{B}}T/\xi_{\mathrm{dom}}^{3} = 0.02 \varepsilon^{(1)}_{T}$. In zero field, $H = 0$, both $\mathrm{d}M/\mathrm{d}H$ and $\mathrm{Re}\,\chi_{\mathrm{ac}}$ assume the value $\langle\chi^{\mathrm{mic}}_{H}\rangle|_{H=0} \approx 0.21$ after zero-field cooling, since $\langle(\hat{H}\hat{Q})^{2}\rangle = 1/3$.

\begin{figure}
\includegraphics[width=1.0\linewidth]{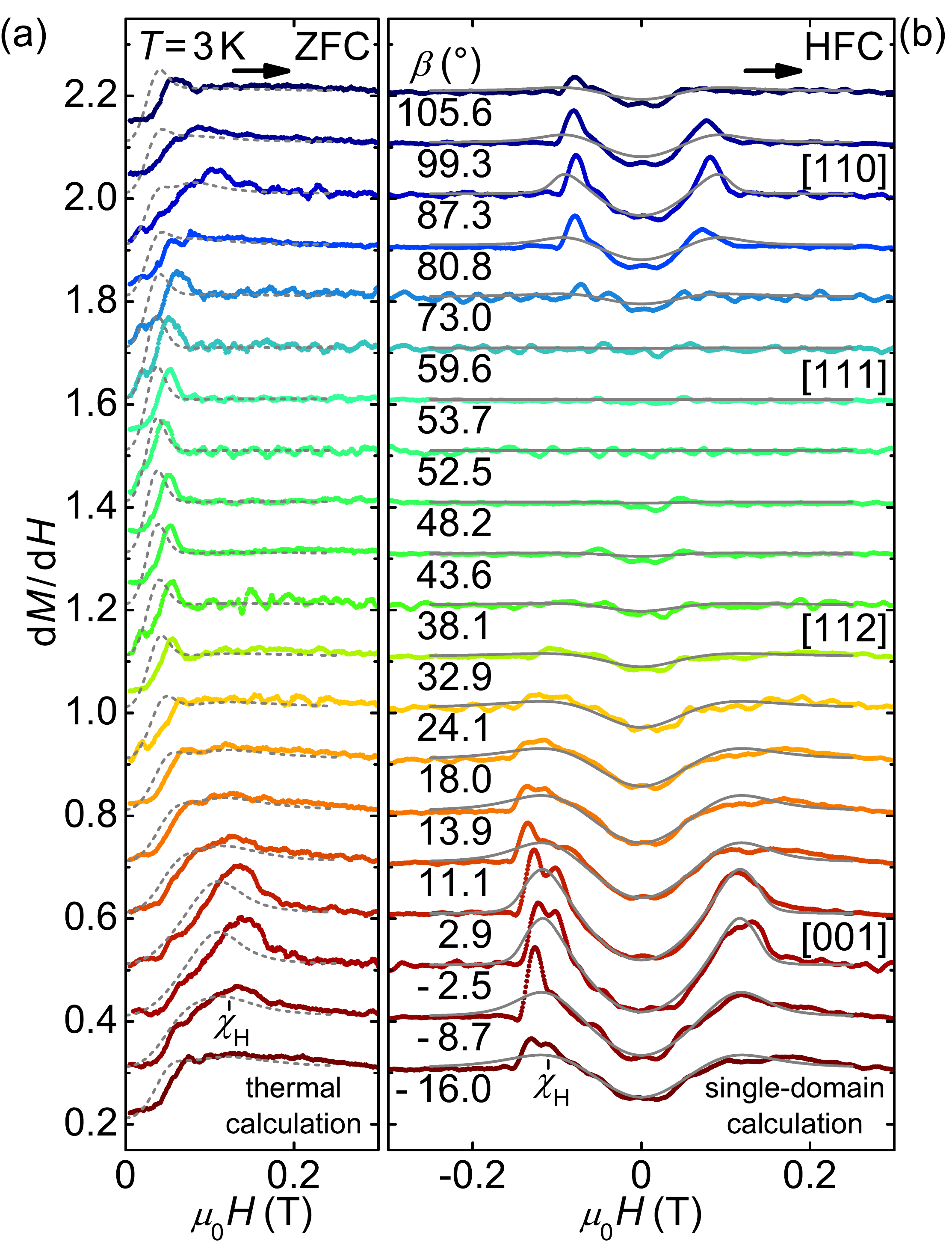}
\caption{Susceptibility calculated from the measured magnetization for a large number of field directions. Magnetic fields $\hat{H}$ are applied along $(\frac{\sin\beta}{\sqrt{2}},\frac{\sin\beta}{\sqrt{2}},\cos\beta)$ parametrized by the angle $\beta$. Data are offset by 0.1 for clarity. (a)~Data measured after zero-field cooling (ZFC). The gray dashed lines are calculations of $\chi_{H}$ assuming a thermal population of the helical domains with $k_{\mathrm{B}}T/\xi_{\mathrm{dom}}^{3} = 0.05\varepsilon^{(1)}_{T}$. (b)~Data measured after high-field cooling (HFC) starting in large negative fields. The gray solid lines are calculations of $\chi_{H}$ assuming a single macroscopic domain.}
\label{figure7}
\end{figure}

In order to demonstrate the very good agreement between experiment and theory, we show the susceptibility calculated from the measured magnetization for a large number of field directions in Fig.~\ref{figure7}. The measurements were carried out on the spherical sample \#1 with applied field directions $\hat{H} = (\frac{\sin\beta}{\sqrt{2}},\frac{\sin\beta}{\sqrt{2}},\cos\beta)$ tracking one of the great circles on the unit sphere shown in gray in Fig.~\ref{figure1}(b). The spherical sample shape ensured that demagnetization effects were unchanged under changes of field direction. In Figs.~\ref{figure7}(a) and \ref{figure7}(b), data recorded after zero-field cooling and starting in the conical state at large negative fields are compared with calculations assuming a thermal population of domains and a single domain, respectively. In general, the evolution of the susceptibility as a function of the angle $\beta$ is described very well by our calculations (gray lines) \footnote{The weak orientation dependence of the conical-to-field-polarized transition at $H_{c2}$ may also be accounted for by magneto-crystalline anisotropies, see the supplemental material for further details~\cite{supplement}.}, where small but distinct deviations might indicate the importance of contributions that are beyond our mean-field approximation, cf.\ Sec.~\ref{Discuss}.

In Fig.~\ref{figure7}(b), discrepancies between theory and experiment are only observed close to the critical fields for $\vec{H}\parallel[110]$ and $\vec{H}\parallel[001]$. We find that $\mathrm{d}M/\mathrm{d}H$ is larger and the signatures of the transitions are more pronounced and robust compared to the calculations. In particular, a clear double peak structure is resolved around $[001]$ ($\beta = 0^{\circ}$) in the experiment, whereas the maxima of the double Ising transitions are smeared in the theoretical curves already for $\beta \approx 3^{\circ}$. Moreover, the behavior of the susceptibility differs for decreasing and increasing field strength with much weaker critical signatures close to the positive critical fields where multiple domains coalesce into a single domain, consistent with the hysteresis observed in neutron scattering. After zero-field cooling, see Fig.~\ref{figure7}(a), the agreement is comparable to the results shown in Fig.~\ref{figure6}, where deviations of the experimental zero-field value from 0.21 are attributed to a systematic error arising from the smoothing algorithm.


\section{Discussion}
\label{Discuss}

We have presented an effective mean-field theory for the helix orientation in the limit of weak magneto-crystalline anisotropies taking into account the symmetries of the tetrahedral point group $T$ of the cubic chiral magnets. With only few phenomenological parameters, this model allowed us to successfully describe the response of the helix pitch vector $\vec{Q}$ to an applied magnetic field $\vec{H}$ in MnSi, as probed by means of neutron scattering, magnetisation and ac susceptibility measurements.

While the overall quantitative agreement is very good, there are, nevertheless, small but systematic discrepancies between theory and experiment close to the critical values of magnetic fields pointing along high-symmetry directions. Most remarkably, our experiments reveal hysteretic behavior at the nominally continuous elastic Ising transitions. Sharp critical signatures are observed when the transition is approached from the single-domain conical state at large fields in both neutron scattering and susceptibility. The signatures, in fact, are even sharper and more robust with respect to deviations of the field direction than theoretically expected. In contrast, comparatively broad features arise when multiple helical domains coalesce, for instance, after zero-field cooling.

We relate these observations to the need for corrections to our mean-field treatment, that were not taken into account, such as thermal fluctuations and topological defects of helimagnetic order. While the former may cause the enhancement of critical signatures, the latter may be the origin of the hysteresis. In fact, the discrepancy between decreasing and increasing field strength close to the Ising transitions is likely to be a non-equilibrium phenomenon. From the analysis of the ac susceptibility we deduced that the reorientation of the pitch vector, $\hat{Q}$, in general possesses a very large relaxation time $\tau_{\hat{Q}}$. We speculate that close to the Ising transitions $\tau_{\hat{Q}}$ might increase further and might even exceed the time scales of our nominally thermodynamic measurements thus giving rise to the hysteretic behavior.  

\begin{figure}
\includegraphics[width=\linewidth]{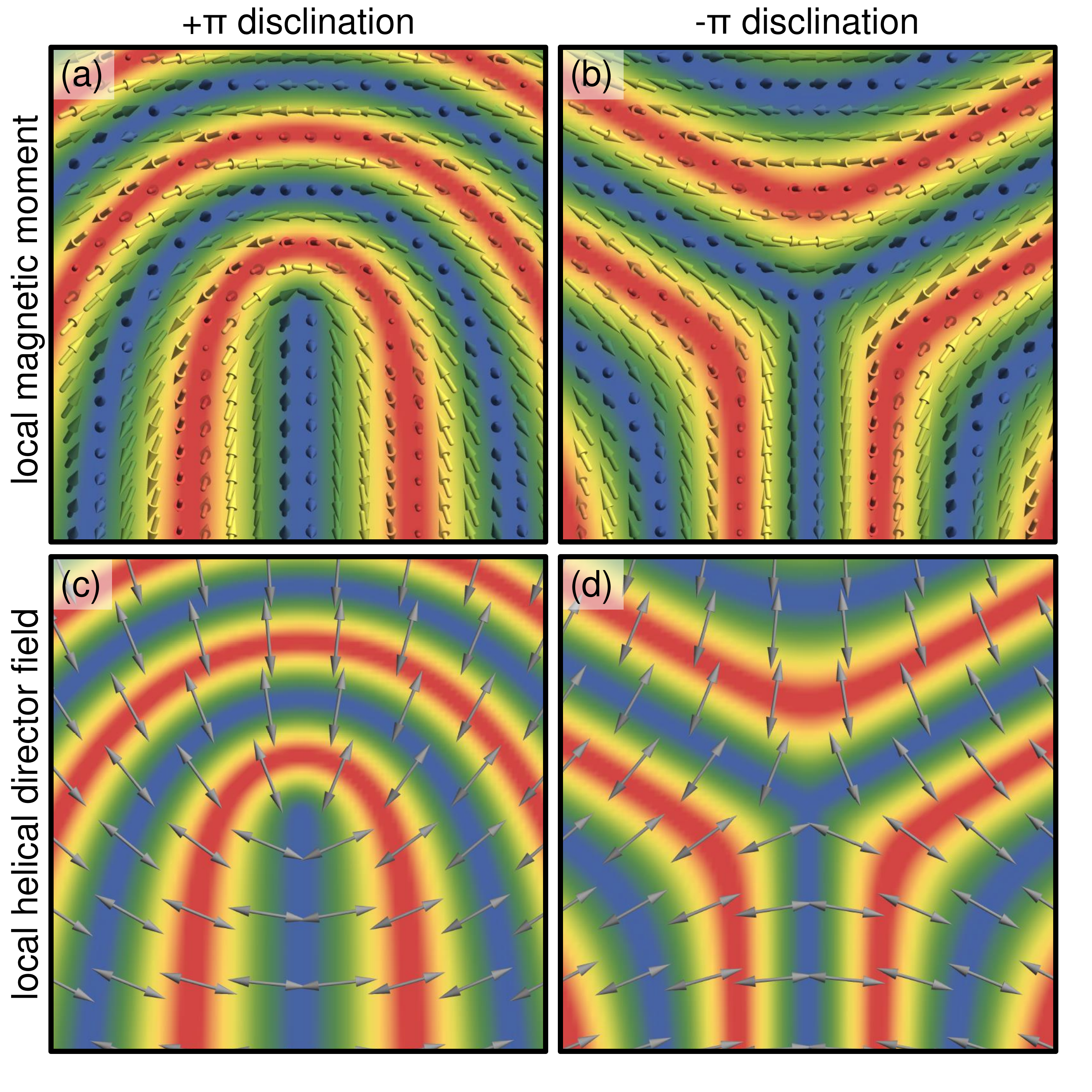}
\caption{Disclinations of helimagnetic order. \mbox{(a),(b)}~Local magnetic moments around a central $+\pi$ and $-\pi$ disclination defect, respectively. The color map encodes the out-of-plane moment. \mbox{(c),(d)}~The local helix axis is represented by a director field (gray double arrows) that rotates by either $+\pi$ or $-\pi$ around the defects. The color maps from the upper panels are shown for comparison.}
\label{figure8}
\end{figure}

A possible origin of a large relaxation time $\tau_{\hat{Q}}$ are domain walls with non-trivial topology~\cite{2012:Li:PhysRevLett} that inhibit the coalescence of domains. In certain cases, the stress in a helix domain wall can be relieved by plastic deformations of the helical arrangement in the form of disclinations, see Fig.~\ref{figure8}(a) and (b), which are known from liquid crystals~\cite{1995:deGennes:Book}. Around such defects, the helix axis, identified up to a sign by the orientation of the vector $\vec{Q}$, rotates by $\pi$ and $-\pi$, respectively. These disclinations thus correspond to vortices in the pitch director field, and, as such, they are topologically protected. As a consequence, two domains separated by a topologically non-trivial domain wall consisting of an arrangement of disclinations cannot be smoothly joined.  The defects must  first be removed from the sample, which results in particularly slow relaxation processes, especially, if they also get pinned by disorder. 

While skyrmions have received a lot of interest recently, see for instance Ref.~\onlinecite{2013:Nagaosa:NatureNano} for a review, the properties of topologically non-trivial disclination defects in chiral magnets remain largely unknown and deserve further study. Generally, we expect that the slow dynamics of disclinations should dominate the equilibration of the magnetization at small fields. In fact, a recent study identified the motion of disclination pairs, i.e., edge dislocations, as the origin for the slow relaxation dynamics of helimagnetic order in FeGe~\cite{2016:Dussaux:NatCommun}. Furthermore, like skyrmion textures, these disclinations are expected to couple efficiently to spin currents so that they should also give rise to interesting spintronic phenomena in helimagnets.

In summary, the helix reorientation transition in the cubic chiral magnets is an elastic transition~\cite{2015:Zacharias:EurPhysJSpecialTopics, 2015:Zacharias:PhysRevLett} that is distinct from conventional phase transitions in magnets. In MnSi the reorientation of the helix pitch vector $\vec{Q}$ as a function of magnetic field $H$ for $\vec{H}\parallel\langle100\rangle$ involves two elastic Ising transitions breaking subsequently a $\mathds{Z}_{2} \times \mathds{Z}_{2}$ symmetry. A single elastic Ising transition occurs for field orientations $\vec{H}\parallel\langle hk0 \rangle$ with $h,k \neq 0$, while for other field directions only a crossover phenomena remains. Helical domains unfavorably populated after zero-field cooling may be depopulated discontinuously for increasing fields. Moreover, slow relaxation processes associated with the helix pitch orientation $\hat{Q}$ quantitatively explain the discrepancy between the susceptibilities $\mathrm{d}M/\mathrm{d}H$ and $\mathrm{Re}\,\chi_{\mathrm{ac}}$. Finally, the hysteretic behavior observed close to the continuous elastic Ising transitions is attributed to a substantial enhancement of relaxation times, presumably due to topologically non-trivial disclination defects.

Our theory is also applicable to other cubic chiral magnets. Interestingly, an orientation of $\hat{Q}$ along $\langle100\rangle$ is favored by the magneto-crystalline anisotropies in Cu$_{2}$OSeO$_{3}$~\cite{2012:Adams:PhysRevLett} and in FeGe close to its critical temperature~\cite{1989:Lebech:JPhysCondensMatter}, implying $\varepsilon_{T}^{(1)} < 0$ in Eq.~\eqref{TPot} in contrast to MnSi. A special situation then arises for $\vec{H}\parallel\langle111\rangle$, where all three helimagnetic $\langle100\rangle$ domains are energetically degenerate. In this case, we predict that the threefold rotation symmetry around $\langle111\rangle$ of the point group $T$ protects an elastic $\mathds{Z}_{3}$ three-state clock transition for the orientation of the helix pitch vector $\vec{Q}$.


\begin{acknowledgments}
We wish to thank T.\ Adams, S.\ Mayr, M.\ Meven, and F.\ Rucker for fruitful discussions and assistance with the experiments. Financial support through DFG TRR80 (From Electronic Correlations to Functionality), DFG FOR960 (Quantum Phase Transitions), DFG SFB1143 (Correlated Magnetism: From Frustration To Topology), and ERC Advanced Grant 291079 (TOPFIT) is gratefully acknowledged. A.B., A.C., M.W., and M.H.\ acknowledge financial support through the TUM graduate school.
\end{acknowledgments}

%


\onecolumngrid
\clearpage

\begin{center}
\textbf{\large Supplementary Material for: Symmetry breaking, slow relaxation dynamics, and topological defects at the field-induced helix reorientation in MnSi}
\end{center}

\setcounter{equation}{0}
\setcounter{figure}{0}
\setcounter{table}{0}
\setcounter{section}{0}
\setcounter{page}{1}
\makeatletter
\renewcommand{\theequation}{S\arabic{equation}}
\renewcommand{\thefigure}{S\arabic{figure}}
\renewcommand{\thesection}{S\Roman{section}}
\renewcommand{\bibnumfmt}[1]{[S#1]}
\renewcommand{\citenumfont}[1]{S#1}


In this supplement, we present additional details of the susceptibility that follow from our theoretical model. We compare the experimentally observed behavior with the predictions as derived for a macroscopic single-domain and a thermal ensemble of domains. Further, susceptibility data are shown covering a wider field range than in the main text, in particular including the second critical field $H_{c2}$. Details of the analysis of our small-angle neutron scattering data are also presented.

\section{Magnetization and Susceptibility}

\subsection{Orientational dependence of the susceptibility}

In the following, we continue the theoretical discussion of Sec.~III\,B of the main text on the implications of the low symmetry of the point group $T$ for the helix reorientation. The reorientation process is symmetry-equivalent for field orientations $[hkl]$, $[klh]$, and $[lhk]$, which follows from the three-fold rotation symmetry around the $\langle111\rangle$ axis of $T$ as illustrated in Fig.~3 of the main text. It differs, however, for magnetic fields along $[hkl]$ and $[khl]$. 

\begin{figure}[b]
\includegraphics[width=0.6\linewidth]{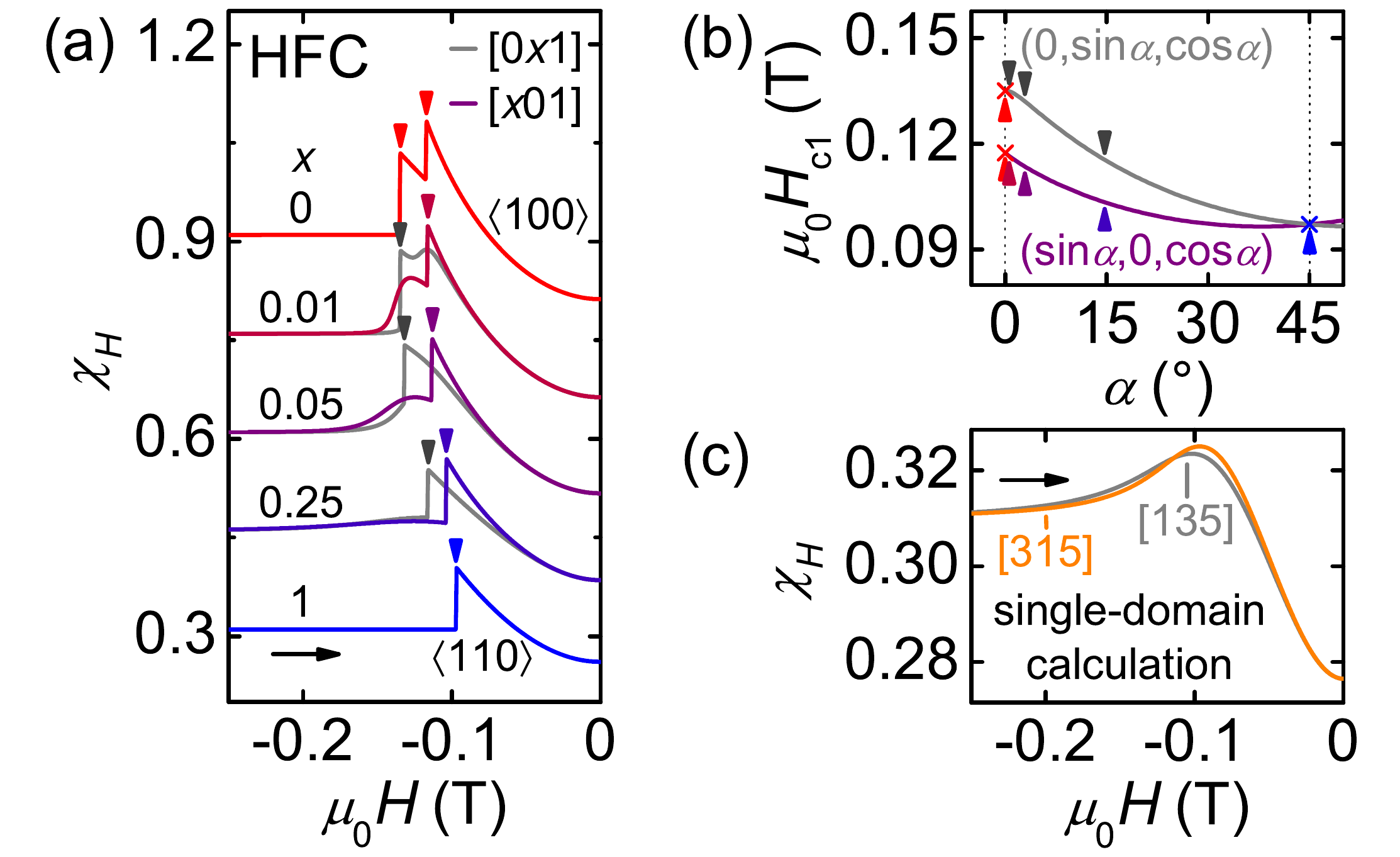}
\caption{(Color online) Theoretical computations of the susceptibility $\chi_{H}$ illustrating the dependence of the helix reorientation process as a function of field direction. (a)~Susceptibility as a function of field for field directions $[0x1]$ and $[x01]$ for various $x$. Data have been offset for clarity. (b)~Critical fields as a function of the applied field direction. Arrows mark the field directions depicted in panel (a). (c)~Susceptibility as a function of field for generic field directions where the helical pitch vector smoothly reorients resulting in a crossover. Field orientations $[135]$ and $[315]$ display slightly different behavior.}
\label{figureS1}
\end{figure}

In particular, the critical field $H_{c1}$ displays distinct behavior along $[0x1]$ and $[x01]$ for $1 > x > 0$. This is illustrated in Fig.~\ref{figureS1}(a), where we show the susceptibility $\chi_{H}$ evaluated according to Eq.~(6) of the main text. Two sharp critical signatures for $x = 0$ are observed, reflecting the two Ising transitions of the helix wavevector for $[001]$ at the two critical fields $H_{c1,>}^{[001]}$ and $H_{c1,<}^{[001]}$ (red arrows). For any finite $x > 0$, i.e., a small tilt away from $[001]$, one of the Ising symmetries is explicitly broken and the corresponding sharp signature in $\chi_{H}$ is rounded, notably either the one at the upper or the lower critical field depending on the field orientation. As a result, the field directions $[0x1]$ and $[x01]$ exhibit distinctly different susceptibility curves and different values of critical fields for the Ising transition which is unchanged in either case. Only for $x = 1$, the same susceptibility is recovered as a function of $H$ and the critical fields coincide for both the $[011]$ and $[101]$ direction. The corresponding evolution of the critical fields is shown in Fig.~\ref{figureS1}(b). 

It is also interesting to note that similar considerations also apply for fields along the generic directions $[hkl]$ ($h,k,l \neq 0$), as exemplified by $[135]$ and $[315]$ shown in Fig.~\ref{figureS1}(c). For this field directions the reorientation process of the helix represents a crossover resulting in a broad maximum in the susceptibility. While qualitatively similar behavior is observed in both cases, the exact position and shape of the maximum slightly differs.

\subsection{Thermal domain ensemble vs.\ macroscopic domains}

\begin{figure}
\includegraphics[width=0.6\linewidth]{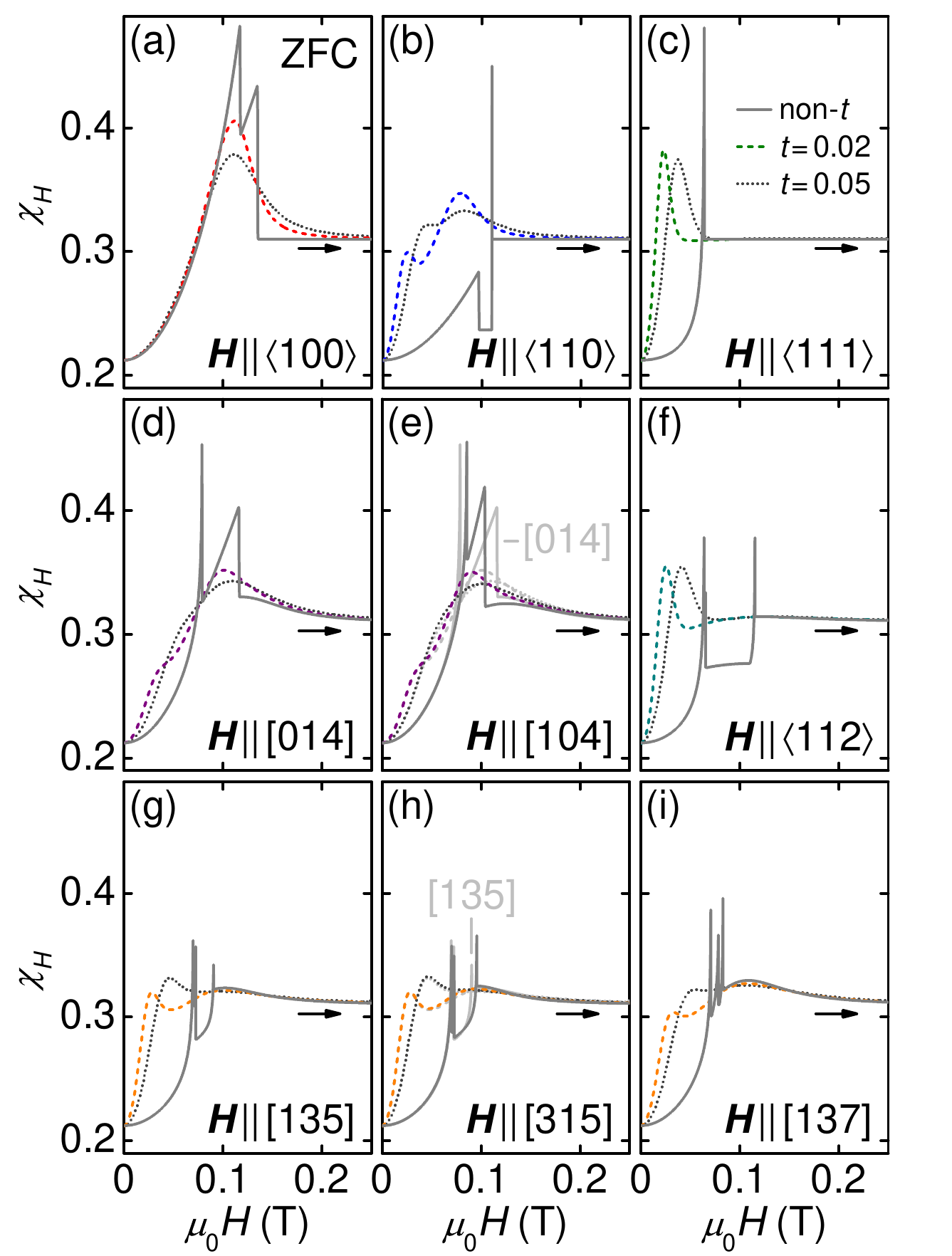}
\caption{(Color online) Theoretically expected behavior of the susceptibility $\chi_{H}$ as a function of field $H$ for different orientations of $\vec{H}$. At zero field, $H = 0$, all four domains are assumed to be populated equally, corresponding to the situation after zero-field cooling. Solid lines assume macroscopic domains that switch at their spinodal point resulting in first-order spikes in $\chi_{H}$. Dashed and dotted lines correspond to a thermal ensemble of domains for two different dimensionless temperatures $t$, see text for details.}
\label{figureS2}
\end{figure}

After zero-field cooling (ZFC) in MnSi, four helix domains with wavevectors $\vec{Q}$ pointing along $[111]$, $[\bar{1}11]$, $[1\bar{1}1]$ and $[\bar{1}\bar{1}1]$ are equally populated. Upon application of a magnetic field, some of these domains are expected to jump with their $\vec{Q}$ towards the field direction in a first-order transition. The details of these first-order transitions depend on the domain dynamics. In the main text, we discussed a simple model assuming a thermal distribution of helix domains with a Boltzmann weight determined by the potential $\mathcal{V}(\hat{Q})$, see Eq.~(7) of the main text. 

For completeness because being instructive, we would like to discuss in the following a different scenario where each domain is macroscopically large so that it is trapped in its local minimum of $\mathcal{V}(\hat{Q})$. As a function of field, metastable local minima eventually become locally unstable at their spinodal points where the domains suddenly switch to the more stable configuration. In Fig.~\ref{figureS2}, we show the resulting susceptibility (solid lines) after ZFC for different field orientations. For comparison, the thermal evaluation according to Eq.~(7) of the main text is also shown as dashed and dotted lines for two dimensionless temperatures $t = k_{\mathrm{B}}T/(\epsilon_{T}^{(1)}\xi_{\mathrm{dom}}^{3})$ with the linear domain size $\xi_{\mathrm{dom}}$. Whereas the susceptibility of the thermal ensemble is always smooth, critical behavior is observed for the macroscopic domains in case of a field along $\langle001\rangle$ as shown in Fig.~\ref{figureS2}(a). Here, all four domains contribute to the critical signatures of both second-order Ising transitions. For the other field directions shown in Fig.~\ref{figureS2}, one or more domains become metastable at a finite applied field. After reaching their respective spinodal point, domains switch leading to first-order spikes in the susceptibility.

The very sharp spikes predicted theoretically are not observed experimentally. Furthermore, comparison with neutron scattering data as well as susceptibility measurements indicate that metastable domains do not reconstruct at their spinodal point but already at a much lower fields, see for example Fig.~4(a) of the main text. We therefore conclude that macroscopic metastable domains 
do not correctly describe the behavior after zero-field cooling. In contrast, stable macroscopic helix domains describe satisfactorily the behavior after field-cooling as explained in the main text.

\subsection{Orientational dependence of $H_{c2}$}

\begin{figure}
\includegraphics[width=0.6\linewidth]{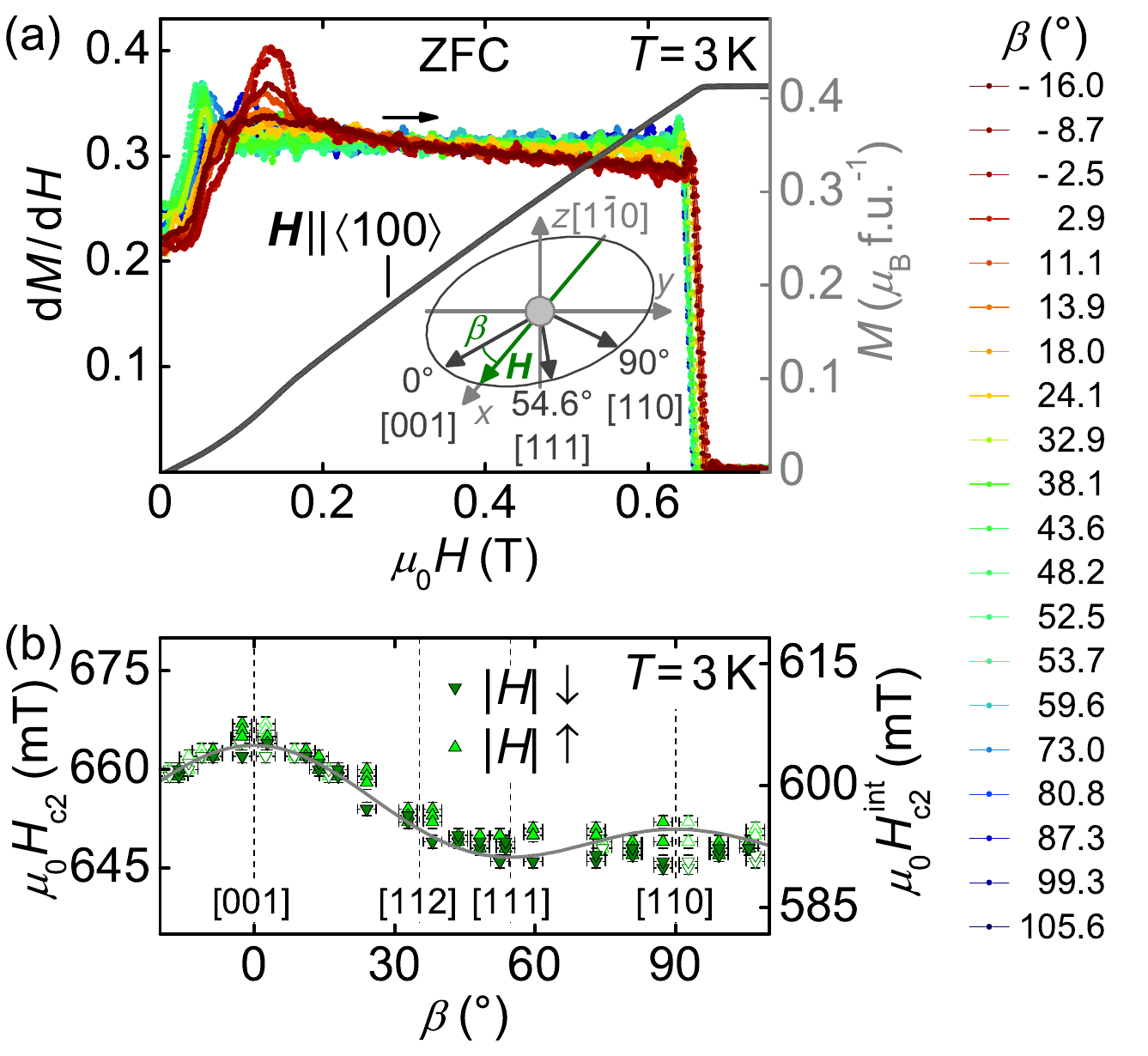}
\caption{(Color online) Orientational dependence of the magnetic properties of MnSi at low temperatures. (a)~Field dependence of the susceptibility calculated from the measured magnetization, $\mathrm{d}M/\mathrm{d}H$, for different orientations of the applied magnetic field as parametrized by $\hat{H} = (\frac{\sin\beta}{\sqrt{2}}, \frac{\sin\beta}{\sqrt{2}}, \cos\beta)$. The sketch illustrates the geometry of the experiment. We also show the measured magnetization for field parallel to $\langle100\rangle$ (gray curve). (b)~Orientational dependence of the second critical field $H_{c2}$ for the transition between the conical and the field-polarized state at low temperatures. The theoretical fit of Eq.~\eqref{Hc2Fit} (gray curve) perfectly accounts for the experimental data (green triangles).}
\label{figureS3}
\end{figure}

In Fig.~\ref{figureS3}(a), we present again the susceptibility $\mathrm{d}M/\mathrm{d}H$ of Fig.~7 of the main text as calculated from the measured magnetization but now for a wider field range. This range includes, in particular, the second critical field $H_{c2}$ where the conical helix disappears at a second-order transition, the magnetization gets field-polarized, and the susceptibility drops accordingly. The data are shown for various field orientations $\hat{H} = (\frac{\sin\beta}{\sqrt{2}}, \frac{\sin\beta}{\sqrt{2}}, \cos\beta)$ parametrized by $\beta$ corresponding to a rotation of the field around the $[1\bar{1}0]$ axis, see sketch in Fig.~\ref{figureS3}(a). We observe that between $H_{c1}$ and $H_{c2}$, the susceptibility remains basically constant except for orientations with small $\beta$ where it decreases slightly.

Shown in Fig.~\ref{figureS3}(b) is the orientational dependence of the second critical field. We define here $H_{c2}$ at the point of inflection where the susceptibility drops from the conical plateau to very small values in the field-polarized state. The largest value of the critical field is found for a field aligned parallel to $[001]$, denoted by $H_{c2}^{[001]}$, and the smallest values is obtained for a field along $[111]$, denoted by $H_{c2}^{[111]}$. The orientational dependence can be described in terms of the lowest cubic invariant of the unit vector $\hat{H}$,
\begin{align} \label{Hc2Fit}
H_{c2} = H_{c2}^{[111]} + \delta H_{c2}\, \frac{3}{2}  \Big(\hat H_{x}^{4} + \hat H_{y}^{4} + \hat H_{z}^{4} - \frac{1}{3}\Big) = 
H_{c2}^{[111]} + \delta H_{c2} \frac{(1 + 3\cos(2\beta))^{2}}{16}
\end{align}
with $\delta H_{c2} = H_{c2}^{[001]} - H_{c2}^{[111]}$. In the last equation, we used the parametrization of $\hat{H}$ in terms of $\beta$. The corresponding fit to the data is shown as a gray solid line. With increasing temperature, the anisotropy $\delta H_{c2}$ decreases and becomes indiscernible at 28~K (not shown). Within the resolution of our experiment, we do not observe any hysteresis around $H_{c2}$ consistent with the expectation for a second-order phase transition.

\section{Small-angle neutron scattering}

\subsection{Experimental setup}

Our small-angle neutron scattering studies were carried out on the spherical sample \#1 (diameter: 5.75~mm) at the diffractometer MIRA2 at the Heinz Maier-Leibnitz Zentrum (MLZ). An incident neutron beam of wavelength $\lambda = 4.5~\textrm{\AA}$ was collimated with two apertures, each $3\times3~\mathrm{mm}^{2}$, located 1.4~m and 0.5~m in front of the sample, respectively. Scattered neutrons were recorded by a CASCADE detector placed 2.6~m behind the sample. This setup results in an azimuthal resolution $\Delta\theta_\mathrm{ins} = 14.5^{\circ}$. The actual azimuthal width originating from the sample was obtained by means of a deconvolution of the instrumental resolution. Low temperatures and the external magnetic field were generated by a closed-cycle cryostat and a set of water-cooled Helmholtz coils, respectively.

\begin{figure}
\includegraphics[width=1.0\linewidth]{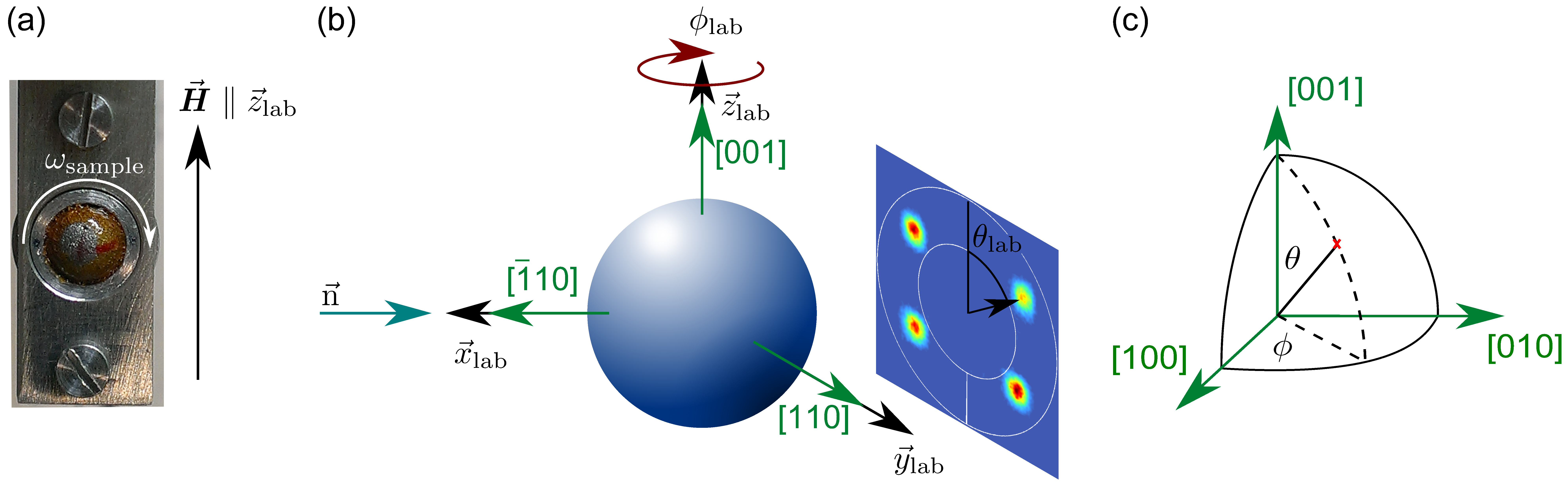}
\caption{(Color online) Setup used for small-angle neutron scattering. (a)~Bespoke sample holder. Rotating the spherical sample by an angle $\omega_{\mathrm{sample}}$ around the axis $\vec(C)_\textup{sample}$ allows to select the crystallographic axis along which the magnetic field is applied. The latter is fixed in vertical direction. (b)~Laboratory and sample coordinate systems indicated by black and green arrows, respectively. The sample (blue sphere) may be rotated by an angle $\phi_\mathrm{lab}$ around $\vec{C}_\mathrm{stick}$. The azimuthal angle in the scattering plane is referred to as $\theta_\mathrm{lab}$. (c)~Definitions of the angles $\theta$ and $\phi$ in the sample coordinate system.}
\label{figureS4}
\end{figure}

As shown in Fig.~\ref{figureS4}(a), using GE varnish, the spherical sample was glued into a bespoke holder that allowed to rotate the sample by $\omega_{\mathrm{sample}}$ around a horizontal axis $\vec(C)_\mathrm{sample}$. With a magnetic field applied along the vertical direction, the latter permitted to choose which crystallographic axis was parallel to the applied field. The sample holder was attached to a sample stick providing the vertical rotation axis $\vec{C}_\mathrm{stick}$ parallel to the external magnetic field. By rotating the sample by $\phi_\mathrm{lab}$ around this axis, different scattering planes may be accessed. The azimuthal angle on the scattering plane was defined as $\theta_\mathrm{lab}$, cf.\ Fig.~\ref{figureS4}(b). The propagation direction of the helices, $\vec{Q}$, may be described in a fixed sample coordinate system. Here, the angles $\phi$ and $\theta$ are the azimuthal and polar angle of spherical coordinates where the equatorial plane is spanned by the crystal axes $[100]$ and $[010]$ as depicted in Fig.~\ref{figureS4}(c).

\begin{figure}
\includegraphics[width=0.6\linewidth]{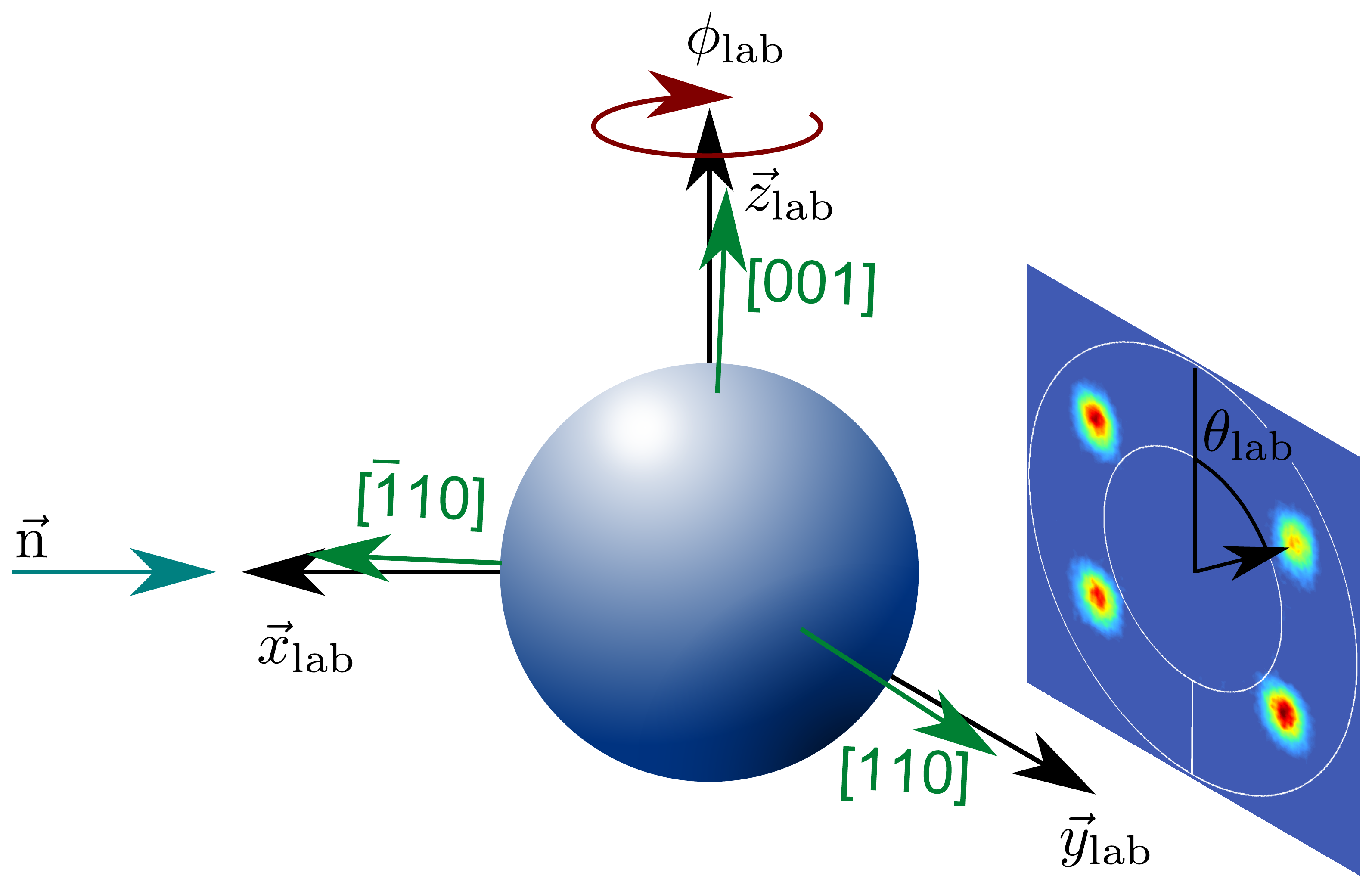}
\caption{(Color online) Illustration of the misalignment of the sample coordinate system with respect to the laboratory frame in a real experiment due to the deviation of the rotation axis $\vec{C}_\mathrm{sample}$ from a crystallographic $\langle110\rangle$ axis.}
\label{figureS5}
\end{figure}

In an ideal experimental setup, $\vec{C}_\mathrm{sample}$ is parallel to a $\langle110\rangle$ axis and thus the external magnetic field may be applied strictly parallel to all major crystallographic axes. As a result, for $\vec{H} \parallel [001]$ the coordinate systems $\left(\phi_\mathrm{lab}, \theta_\mathrm{lab}\right)$ and $\left(\phi, \theta\right)$ are equivalent. In a real experiment, however, a small misalignment between $\vec{C}_\mathrm{sample}$ and the crystalline $\langle110\rangle$ axis may not be avoided. As shown in Fig.~\ref{figureS5}, also the coordinate systems of the sample and the laboratory are tilted with respect to each other. Furthermore, due to the length of the rotatable sample stick, its rotation axis $\vec{C}_\mathrm{stick}$ may not be in line with the center of the sample resulting in a precession of the sample as a function of $\phi_\mathrm{lab}$. As we describe in the following, both deviations may be accounted for in terms of spherical coordinate transformations.

\subsection{Data analysis}

\begin{figure}
\includegraphics[width=1.0\linewidth]{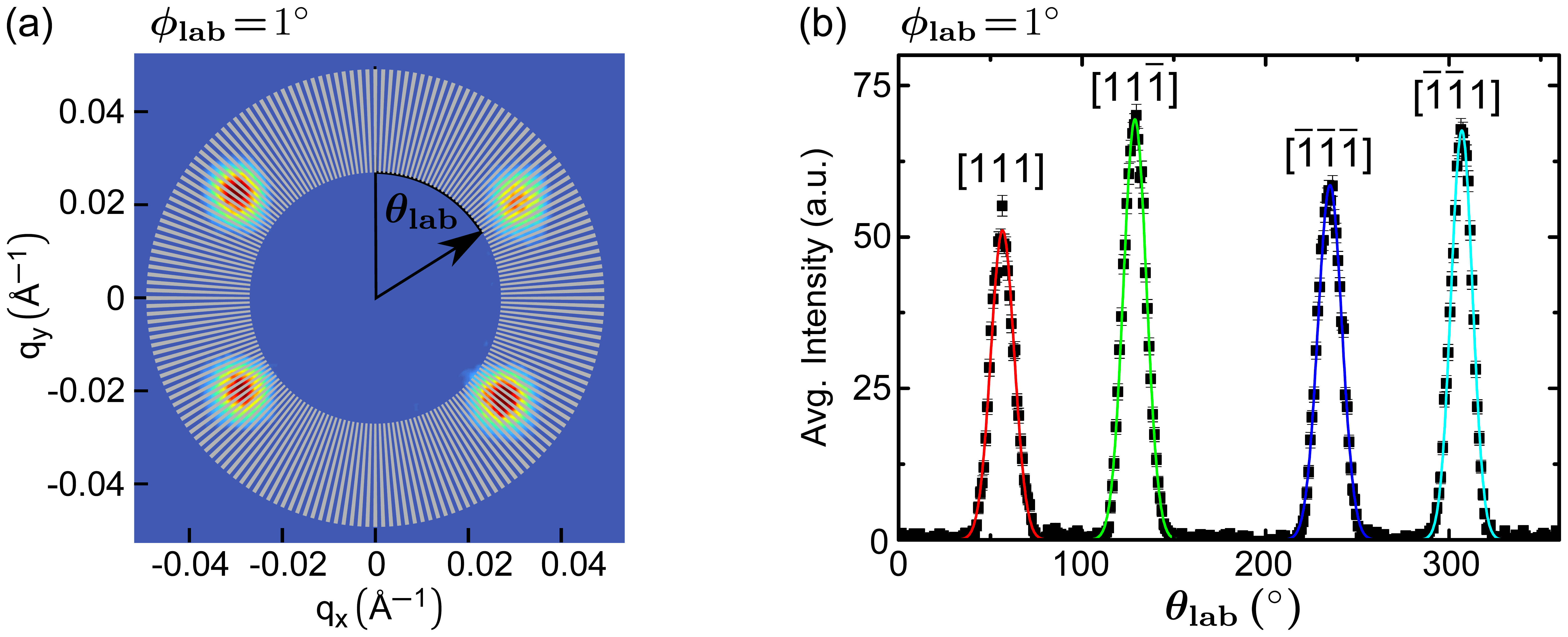}
\caption{(Color online) Determination of the averaged intensity as a function of azimuthal angle $\theta_\mathrm{lab}$. (a)~Typical small-angle neutron scattering pattern. The intensity is averaged over windows of $1^{\circ}$ width in $\theta_\mathrm{lab}$ and $0.022\,\textrm{\AA}^{-1}$ width in $|\vec{Q}|$ around $|\vec{Q}_{c}| = 0.038~\textrm{\AA}^{-1}$. The white regions represent every other window. (b)~Averaged intensity as a function of $\theta_\mathrm{lab}$. Each peak is fitted with a Gauss function and attributed to a given helical domain via $\theta_\mathrm{lab}$ and $\phi_\mathrm{lab}$.}
\label{figureS6}
\end{figure}

For a given temperature and magnetic field value, small-angle scattering patterns were recorded in $1^{\circ}$ steps over a $180^{\circ}$ range in $\phi_\mathrm{lab}$. For the analysis these patterns were associated with polar coordinates where the radius and the azimuthal angle were given by $|\vec{Q}|$ and $\theta_\mathrm{lab}$, respectively. For each pattern we averaged the scattered intensity radially over areas of $1^{\circ}$ width in $\theta_\mathrm{lab}$ and $0.022~\textrm{\AA}^{-1}$ width in $|\vec{Q}|$ around $|\vec{Q}_{c}| = 0.038~\textrm{\AA}^{-1}$ relative to a fixed origin as illustrated in Fig.~\ref{figureS6}(a). In Fig.~\ref{figureS6}(b) we show that the resulting averaged intensity as a function of $\theta_\mathrm{lab}$ exhibits pronounced maxima that are well-described by Gauss functions. The peak positions in $\theta_\mathrm{lab}$ are obtained by means of these fits. However, in the next step the values of $\theta_\mathrm{lab}$ have to be corrected for the tilt and precession of the sample.

\subsubsection{Sample precession}

\begin{figure}
\includegraphics[width=1.0\linewidth]{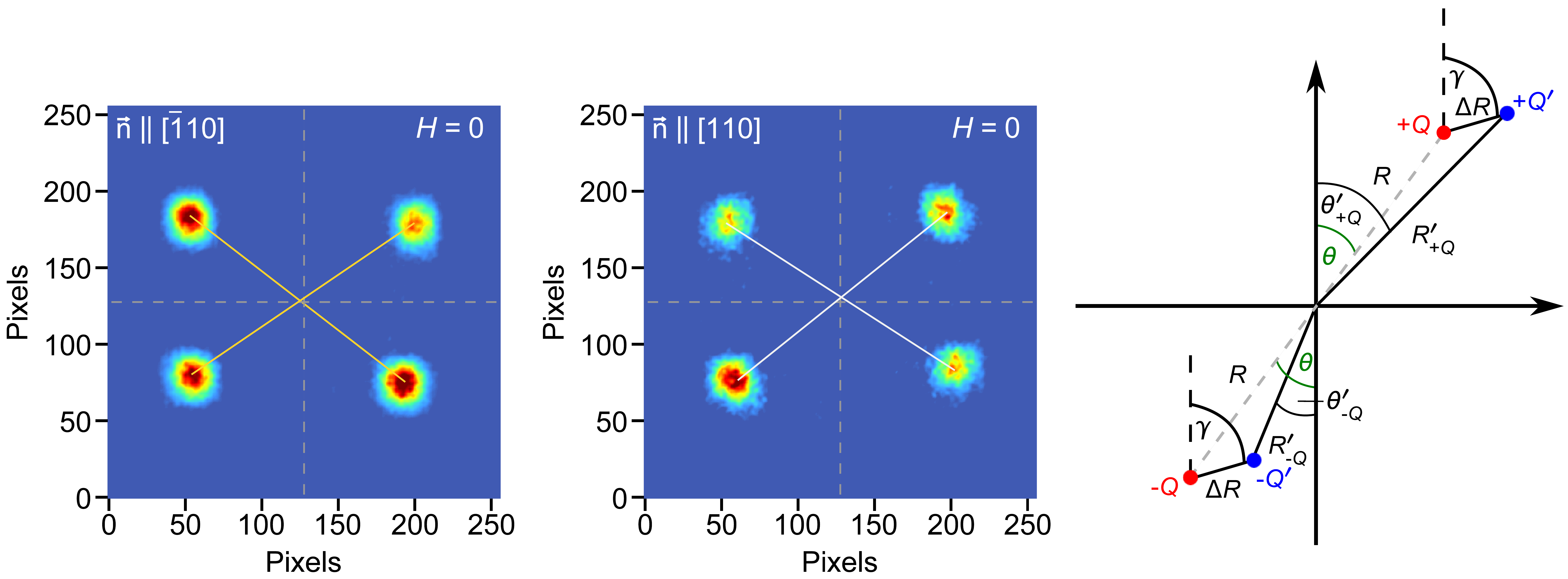}
\caption{(Color online) Effects of the precession of the sample on the scattering patterns. (a)~Pattern for neutron beam $\vec{n}\parallel[\bar{1}10]$. The center of the magnetic Bragg peaks (intersection of yellow lines) is shifted slightly to the left with respect to the fixed origin of the analysis (intersection of gray lines). (b)~Pattern for neutron beam $\vec{n}\parallel[110]$, i.e., after rotating the sample by $90^{\circ}$ in $\phi_\mathrm{lab}$. The center of the magnetic Bragg peaks is shifted to the right of the fixed origin. (c)~Geometrical analysis of the precession. Starting from its intrinsic position (no prime), a pair of Bragg peaks at $\vec{Q}$ and $-\vec{Q}$ is shifted as parametrized by $\Delta R$ and the angle $\gamma$ resulting in the position as obtained from our analysis (prime). See text for details.}
\label{figureS7}
\end{figure}

As mentioned above, the rotation axis of the sample stick $\vec{C}_\mathrm{stick}$ is not in line with the center of the sample resulting in precession of the latter. As illustrated in Figs.~\ref{figureS7}(a) and (b), this movement of the sample causes a small shift of the origin of the scattering plane as a function of $\phi_\mathrm{lab}$. The consequence of the latter on the data analysis, as carried out in the fixed laboratory frame, are sketched in Fig.~\ref{figureS7}(c). Here, quantities with a prime represent the results of the analysis described above, while quantities without a prime correspond to the intrinsic values. The precession of the sample results in small deviations that are parametrized by the shift $\Delta R$ and the angle $\gamma$. As our analysis averages over $|\vec{Q}|$, we disregard $|Q'_{-}|$ and $|Q'_{+}|$ and focus on the angles $\theta'_{-Q}$ and $\theta'_{+Q}$. The arithmetic mean of both angles yields the angle $\bar{\theta'}$ for a given domain. The latter is close to the intrinsic value of $\theta$, as will be discussed by means of geometric considerations in the following.  

Starting from 
\begin{align*}
\theta'_{\textup{+}Q} &= \theta + \alpha\\
\theta'_{\textup{-}Q} &= \theta - \beta  \\
\bar{\theta'} &= \theta + \frac{\alpha - \beta}{2}
\end{align*}
with 
\begin{align*}
\alpha &= \arccos\left(\frac{|\vec{Q}|+\Delta R\cos\,(\gamma-\theta)}{\sqrt{|\vec{Q}|^{2} + \Delta R^{2} + 2|\vec{Q}|\Delta R\cos\,(\gamma-\theta)}}\right)\\
\beta &= \arccos\left(\frac{|\vec{Q}|-\Delta R\cos\,(\gamma-\theta)}{\sqrt{|\vec{Q}|^{2} + \Delta R^{2} - 2|\vec{Q}|\Delta R\cos\,(\gamma-\theta)}}\right),
\end{align*}
we obtain the deviation of $\bar{\theta'}$ from the intrinsic value $\theta$, 
\begin{align*}
\Delta\bar{\theta '}=\bar{\theta'} - \theta = \frac{\alpha-\beta}{2} = \frac{1}{2}\arccos\left(\frac{|\vec{Q}|^{2} - \Delta R^{2}\cos\,(2\gamma - 2\theta)}{\sqrt{|\vec{Q}|^{4} - 2|\vec{Q}|^{2}\Delta R^{2}\cos\,(2\gamma-2\theta)+\Delta R^{4}}}\right).
\end{align*}

The maximum value of $\Delta \bar{\theta '}$ is obtained when the cosine in the expression becomes zero. Thus, the largest deviation as a function of $\Delta R \slash |\vec{Q}|$ is given by 
\begin{align*}
\Delta\bar{\theta'}_\mathrm{max} = \frac{1}{2}\arccos\left(\frac{1}{\sqrt{1+\frac{\Delta R}{|\vec{Q}|}^{4}}}\right).
\end{align*}

\begin{figure}
\includegraphics[width=0.7\linewidth]{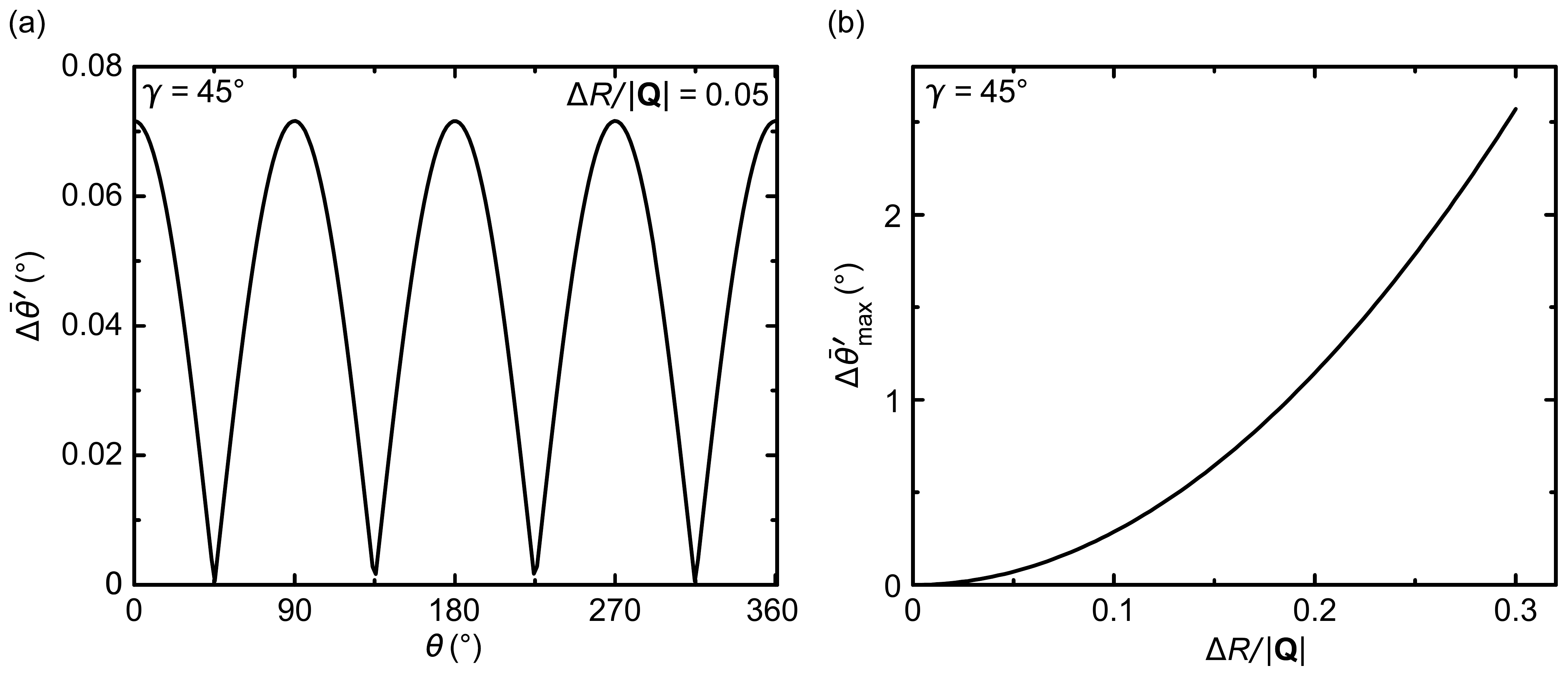}
\caption{(Color online) Effects of the precession of the sample on the extracted domain position. (a)~Deviation of the angle $\bar{\theta'}$ from the intrinsic value $\theta$ as a function of $\theta$ for a relative shift $\delta R \slash |\vec{Q}| = 5\%$ and $\gamma = 45^{\circ}$. (b)~Maximum deviation $\Delta\bar{\theta'}_\mathrm{max}$ as a function of relative shift $\Delta R \slash R$ for $\gamma = 45^{\circ}$.}
\label{figureS8}
\end{figure}

As shown in Fig.~\ref{figureS8}(a), for typical values such as $\Delta R \slash |\vec{Q}| = 5\%$ and $\gamma = 45^{\circ}$, the deviation $\Delta\bar{\theta'}$ remains very small, below $0.1^{\circ}$, for all positions $\theta$. In fact, as illustrated in Fig.~\ref{figureS8}(b), $\Delta\bar{\theta'}$ increases only slowly for small deviations reaching a value of only about $1^{\circ}$ for a relative shift $\Delta R \slash |\vec{Q}| = 20\%$. Note, however, that in our study we observed $\Delta R \slash |\vec{Q}| < 5\%$ for all angles $\phi_\mathrm{lab}$ measured implying only minor consequences on the extracted domain positions due to the precession of the sample. Therefore, the effects of the latter are neglected in our analysis.

\subsubsection{Sample tilt}

The angles discussed in the previous section are defined within the laboratory frame. In order to compare our experimental results with our theoretical calculations, the angles have to be transformed into the internal coordinate system of the sample as defined by the crystallographic $\langle100\rangle$ axes, cf.\ Fig.~\ref{figureS4}(c).

\begin{figure}
\includegraphics[width=0.7\linewidth]{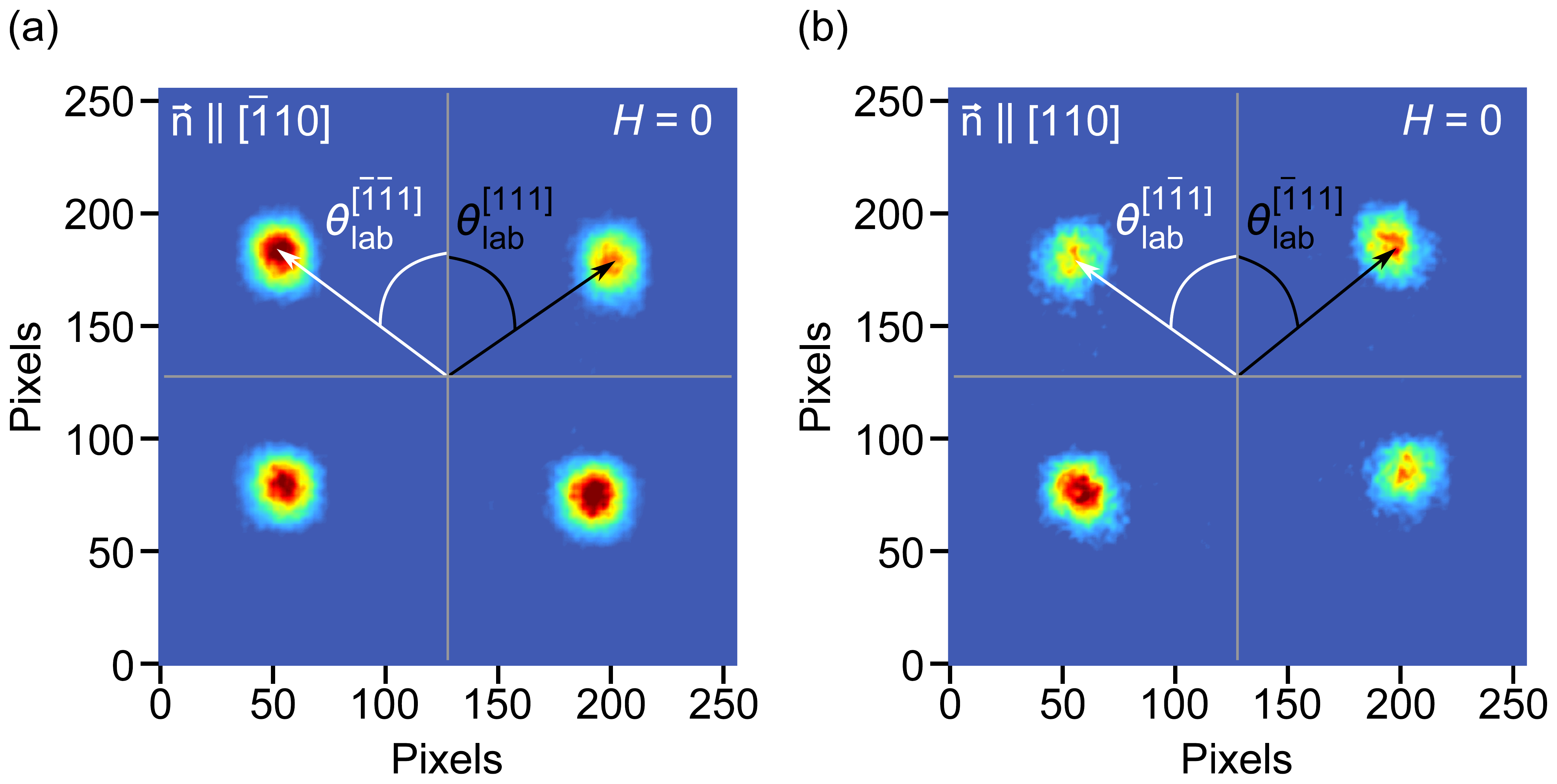}
\caption{(Color online) Determination of the tilt of the sample with respect to the laboratory frame. (a)~Scattering pattern for neutron beam $\vec{n}\parallel[\bar{1}10]$ yielding the tilt angle $\delta_{1}$. (b)~Scattering pattern for neutron beam $\vec{n}\parallel[110]$ yielding the tilt angle $\delta_{2}$. See text for details.}
\label{figureS9}
\end{figure}

The tilt of the coordinate systems of the sample with respect to the laboratory frame may be described in terms of two angles, $\delta_{1}$ and $\delta_{2}$, which account for the misalignment towards the crystallographic $[\bar{1}10]$ and $[110]$ axes, respectively. The angles are defined as
\begin{align*}
\delta_{1} &= \frac{\theta_\mathrm{lab}^{[111]} - \theta_\mathrm{lab}^{[\bar{1}\bar{1}1]}}{2}
\delta_{2} &= \frac{\theta_\mathrm{lab}^{[\bar{1}11]} - \theta_\mathrm{lab}^{[1\bar{1}1]}}{2}.
\end{align*}

Note that the angle $\theta_\mathrm{lab}^{[111]}$ is obtained by averaging over the pair of magnetic Bragg peaks along $[111]$ and $[\bar{1}\bar{1}\bar{1}]$ as measured after zero-field cooling. Figures~\ref{figureS9}(a) and \ref{figureS9}(b) illustrates the corresponding tilt for the planes perpendicular to $[\bar{1}10]$ and $[110]$. An analysis of the scattering patterns yield tilt angles $\delta_{1} = 2.3^{\circ}$ and $\delta_{2} = -2.8^{\circ}$. In contrast to the effects of the precession of the sample discussed in the previous section, these misalignment angles are quite considerable. Hence, after extracting the domain positions $\vec{Q}$ in the laboratory coordinate system, the sample tilt is corrected by the subsequent application of two rotation matrices finally yielding the values of $\theta$ and $\phi$ in the sample coordinate system as presented in the main text. For the analysis of neutron scattering data for $\vec{H} \parallel [110]$, we follow the same procedure.

\subsubsection{Visualization of scattered intensity on the surface of a sphere}

The spheres of scattered intensity shown in the main text are constructed by creating a scalar field of the intensity distribution $I\,(\theta)$ for each value of $\phi_\mathrm{lab}$. Two spherical coordinate transformations are used to correct for the tilting discussed above. As discussed above, two spherical coordinate transformations are used to correct for the tilting of the sample with respect to the laboratory frame, while the influence of the precession of the sample is neglected. Further note that the instrumental resolution was not deconvolved. Due to the latter the intensity maxima on the spheres may appear broader than the corresponding error bars in the other figures of the main text.

\end{document}